\documentclass[%
 reprint,
showpacs,
 amsmath,amssymb,
 aps,
prb,
]{revtex4-1}

\usepackage{graphicx}
\usepackage{dcolumn}
\usepackage{bm}

\newcommand{\cubo}{Cu$_3$(BO$_3$)$_2$} 
\newcommand{\cume}{CuB$_2$O$_4$}       
\newcommand{\cugeo}{CuGeO$_3$}

\bibpunct{[}{]}{;}{n}{}{}

\begin{document}


\title{Lattice dynamics and electronic transitions in a structurally-complex \\ layered copper borate \cubo{}}

\author{A. D. Molchanova}
 \email{nastyamolchanova@list.ru}
 \altaffiliation[Also at the ]{Institute of Spectroscopy, Russian Academy of Sciences, 108840 Moscow, Troitsk, Russia}
\author{M. A. Prosnikov}
\affiliation{Ioffe Physical Technical Institute, Russian Academy of Sciences, 194021 St.-Petersburg, Russia}%
\author{K. N. Boldyrev}
\affiliation{Institute of Spectroscopy, Russian Academy of Sciences, 108840 Moscow, Troitsk, Russia}
\author{R. M. Dubrovin}
\affiliation{Ioffe Physical Technical Institute, Russian Academy of Sciences, 194021 St.-Petersburg, Russia} 
\author{V. Yu. Davydov}
\affiliation{Ioffe Physical Technical Institute, Russian Academy of Sciences, 194021 St.-Petersburg, Russia}
\author{A. N. Smirnov}
\affiliation{Ioffe Physical Technical Institute, Russian Academy of Sciences, 194021 St.-Petersburg, Russia} 
\author{M. N. Popova}
\affiliation{Institute of Spectroscopy, Russian Academy of Sciences, 108840 Moscow, Troitsk, Russia}
\author{R. V. Pisarev}
\affiliation{Ioffe Physical Technical Institute, Russian Academy of Sciences, 194021 St.-Petersburg, Russia}

\date{\today}

\begin{abstract}
Copper borate \cubo{} is a complex compound with a layered crystallographic structure in which the Jahn\,--\,Teller active and magnetic copper Cu$^{2+}$ ions occupy sixteen nonequivalent positions in the unit cell displaying controversial magnetic behavior.
In this paper, we report on the infrared and Raman spectroscopic studies of the lattice dynamics and the electronic structure of 3$d^9$ copper states below the fundamental absorption band. 
The lattice dynamics is characterized by a large number of phonons due to a low \textit{P$\overline{1}$} space group symmetry and a large unit cell with \textit{Z}\,=\,10.
Unusually rich set of phonons was found in the low-energy part of the infrared and Raman spectra below~100\,cm$^{-1}$, which we tentatively assign to interlayer vibrations activated by a crystal superstructure and/or to weak force constants for modes related to some structural groups.
Several phonons show anomalous behavior in the vicinity of the magnetic phase transition at~\textit{T}$_N$=10\,K thus evidencing pronounced magnetoelastic interaction.
No new phonons were found below~\textit{T}$_N$, which excludes the spin\,--\,Peierls type of the magnetic transition.
In the region of electronic transitions, a strong broad absorption band centered at $\sim$1.8\,eV is observed, which we assign to overlapping of transitions between the 3$d^9$ states of Cu$^{2+}$ ions split by the crystal field in nonequivalent positions. 
The fundamental charge-transfer absorption band edge has a complex structure and is positioned around $\sim$2.8--3.0\,eV.   


\end{abstract}

\maketitle


\section{INTRODUCTION}

Numerous copper oxide compounds display a wide variety of unique chemical and physical properties and find important applications.
Many of these properties are related, first of all, to an intrinsic ability of copper ions to adopt different valence states, typically Cu$^{1+}$, Cu$^{2+}$, Cu$^{3+}$, and others.
Another intrinsic property of copper ions is their ability to occupy crystallographic positions with different coordination that results, as a rule,  in markedly-distorted geometries.
There are many compounds in which copper ions occupy positions with~2, 3, 5, and 6 nearest-neighbor anions.
In some cases, these specific properties of copper ions are consequences of the electronic Jahn\,--\,Teller effect~\cite{bersuker2013jahn}.
For example, Cu$^{2+}$ ions with the 3\textit{d}$^9$ ionic state are susceptible to distortions in octahedral sixfold-coordinated sites.
Copper ions play a decisive role in many materials by creating their specific physical properties and some examples can be cited.
First of all, that is the simplest, from the point of view of chemical composition, antiferromagnet CuO which demonstrates multiferroic properties~\cite{kimura2008cupric}.
Another simple compound is a diamagnetic Cu$_{2}$O in which the existence of giant Rydberg excitons with principal quantum numbers as large as \textit{n}\,=\,25 has recently been demonstrated \cite{kazimierczuk2014giant}.
A singlet magnetic ground state is realized due to a spatial spin-dimerization of spins \textit{S}\,=\,1/2 at the spin\,--\,Peierls phase transition in a chemically-simple quasi-one-dimensional compound \cugeo{}~\cite{hase1993observation}.
SrCu$_{2}$O$_{3}$ is an example of a two-leg spin-ladder system~\cite{azuma1994observation}.
Strontium-boron cuprate SrCu$_{2}$B$_{2}$O$_{6}$ is a rare example of a quasi-two-dimensional crystal with a singlet magnetic ground state~\cite{kageyama1999exact, kageyama2000direct, kodama2002magnetic}.
A complex mineral herbertsmithite Cu$_{3}$Zn(OH)$_{6}$Cl$_{2}$ is a spin \textit{S}\,=\,1/2 kagome-type spin-liquid system~\cite{bert2007low}.
Multiferroic properties~\cite{park2007ferroelectricity} and anomalous optical properties~\cite{pisarev2006anomalous} were reported in a mixed-valence antiferromagnet LiCu$_{2}$O$_{2}$ in which Cu$^{1+}$ and Cu$^{2+}$ ions coexist.
Another complex-structure example is trigonal green dioptase Cu$_6$[Si$_6$O$_{18}$]$\cdot$6H$_2$O.
Neutron scattering experiments were done on natural crystals, revealing complex magnetic ground state and excitation spectra~\cite{podlesnyak2016coupled}.
Other noticeable and well-studied chiral magnetic cuprate is Cu$_2$OSeO$_3$, known as a skyrmion and half-skyrmion host~\cite{janson2014skyrmions}.
Besides that, there is a lot of examples of molecular magnets based on Cu$^{2+}$~\cite{landee2013recent}.
Many other examples can be given.
And last but not least, high-\textit{T}$_C$ copper-oxide superconductors are, without any doubt, among the most interesting, important, and actively studied materials~\cite{keimer2015quantum}.

Among numerous oxide cuprates, there are several compounds in the binary system CuO-B$_{2}$O$_{3}$.
The simplest compound in this group is a very interesting copper-boron oxide \cume{}, which crystallizes in the space group~${I\overline{4}2d}$ (\#122, \textit{Z}\,=\,12)~\cite{martinez1971crystal}.
This crystal possesses a very intricate magnetic phase diagram with commensurate and incommensurate magnetic structures, multiple magnetic phase transitions, and a separate ordering of the two magnetic Cu$^{2+}$-subsystems~\cite{boehm2003complex}.
Linear and nonlinear optical studies in the region of transitions within the 3$d^9$ configuration split by the crystal field made it possible to observe a unique of set all zero-phonon~(ZP) lines accompanied by a rich picture of phonon sidebands~\cite{pisarev2004magnetic, pisarev2011electronic, pisarev2013lattice}.
Among other results, this material provided a rare, and probably a unique, possibility to evaluate the genuine crystal-field parameters using the exact positions of the ZP lines for the both Cu$^{2+}$  subsystems~\cite{pisarev2011electronic}.
Moreover, ZP~lines allowed one to observe the antiferromagnetic dichroism and revealed new details in the magnetic phase diagram~\cite{boldyrev2015antiferromagnetic}.

These observations of intriguing magnetic properties and unique manifestations of the fine structure in the electronic spectra of \cume{} raise the question to what degree they are related either to its particular crystal structure or to the chemical composition of an oxide compound which contains only Cu$^{2+}$ and B$^{3+}$ cations~\cite{[{It is interesting to note that Pd(4$d^8$)B$_{2}$O$_{4}$, according to }] [{ is the only other known crystal that crystallizes in the same~$I\overline{4}2d$ group. However, to the best of our knowledge no reports on its magnetic and optical properties is available in literature.}] depmeier1982palladium}. This question remains unanswered so far.

Also other copper borates belong to a binary family 3CuO$\cdot$B$_{2}$O$_{3}$.
Unlike \cume{} with a unique structure, they can crystallize in a triclinic~\cite{behm1982pentadecacopper}, a monoclinic~\cite{kuratieva2009new}, and an orthorhombic~\cite{zhang2017synthesis} structures. 
To the best of our knowledge, there are no \textit{ab initio} calculations which could explain structural stability or metastability of these three phases.
We may suppose that in the case when chemically the same compound can adopt different crystallographic structures is a good example of a large versatility of copper ions for occupying crystallographically different positions.
In this group of three materials, the triclinic  \cubo{}~ compound is the most actively studied one \cite{petrakovskii1999synthesis, petrakovskiui2007magnetic, kudo2003anisotropic, kudo2001antiferromagnetic, balaev2016interrelation, sakurai2002antiferromagnetic, fukaya2001long}, practically nothing is known about physical properties of the two other \cubo{} compounds.

However, up to now, no data are available in literature on the lattice dynamics and electronic structure of this crystal. In the present paper, we report on a detailed study of the lattice dynamics, electronic \textit{d-d} transitions, and optical properties of this triclinic crystal, which, from the chemical point of view, is similar to \cume{}.
Both these compounds belong to the binary system CuO-B$_{2}$O$_{3}$, but the former has the composition 3CuO$\cdot$B$_{2}$O$_{3}$ whereas the latter CuO$\cdot$B$_{2}$O$_{3}$.
Obviously, these two compounds possess different crystallographic structures, with the space groups $P\overline{1}$ (\textit{Z}\,=\,10) and  $I\overline{4}2d$ (\textit{Z}\,=\,12), respectively.
The unit cell volume and the number of inequivalent Cu$^{2+}$ positions in \cubo{} is much larger than those in \cume{}.
Both \cubo{} and \cume{} were reported to be good photocatalysts~\cite{liu2013visible}, but no optical data on the former compound are available so far.
We will show that 3\textit{d} electronic spectra of these two compounds strongly differ but the main transitions are observed in the same spectral range.
Possible reasons of similarities and differences between optical properties of these crystals will be discussed. 

The paper is organized as follows.
In Sec.~\ref{sec:crystal_structure}, the crystal structure and other properties of \cubo{} are briefly reviewed.
In Sec.~\ref{sec:exp_details}, experimental details of the infrared, Raman, and optical absorption measurements are described. 
In Sec.~\ref{sec:lattice_dynamics}, we analyze the phonon modes and discuss experimental results of infrared and Raman studies.
In Sec.~\ref{sec:optical_properties}, we analyze the local positions of Cu$^{2+}$ ions and the relevant electronic structure, describe the optical properties and discuss them in comparison with literature data on some other cuprate oxides.
Sec.~\ref{sec:disscussion} deals with conclusions.

\section{CRYSTAL STRUCTURE AND MAGNETIC PROPERTIES}\label{sec:crystal_structure}
\subsection{Crystal structure}
\begin{figure}
\centering
\includegraphics[width=\columnwidth]{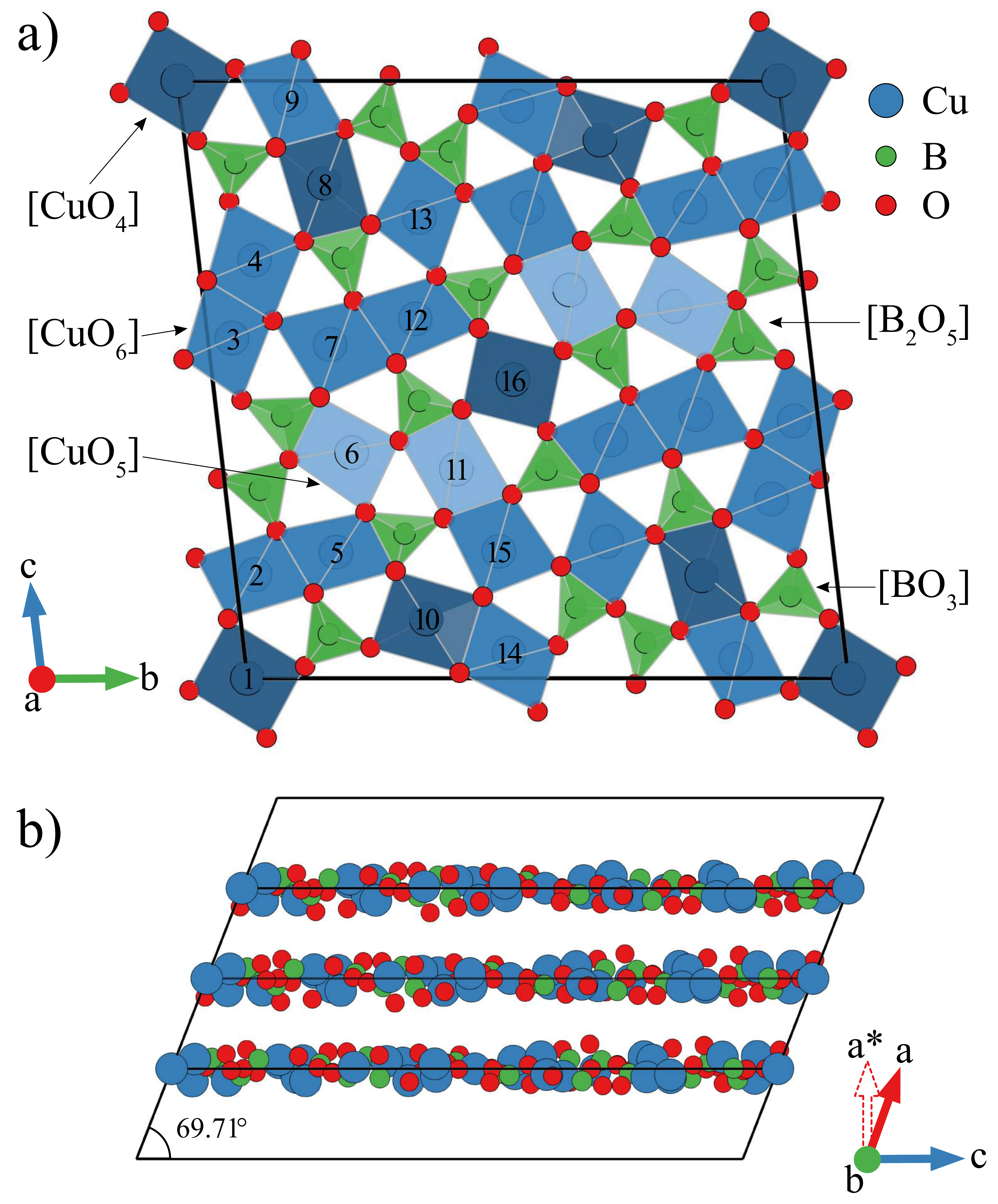}
\caption{\label{fig:structure}
(Color Online) a) Polyhedral representation  of the crystal structure projected along the [100] direction.
Numbers correspond to nonequivalent crystallographic positions of the Cu$^{2+}$ ions.
Distorted [CuO$_6$] groups are shown in blue, near-square coordinated groups by dark-blue, and pyramidal groups by light-blue color. 
b) Projection along the [010] direction; set of the crystal layers in the $bc$ plane with perfect cleavage. Note that the $a$* axis is normal to the cleavage layers.
Single-crystal unit cell is marked by thin solid lines.}
\end{figure}

The triclinic crystal structure of \cubo{} with the lattice parameters \textit{a}\,=\,3.353(2), \textit{b}\,=\,19.665(7), \textit{c}\,=\,19.627(8)\,\AA{}, $\alpha$\,=\,88.77$^{\circ}$, $\beta$\,=\,69.71$^{\circ}$, and $\gamma$\,=\,69.24$^{\circ}$ was solved in Ref.~\cite{[{}] [{ \\ Structural data (.cif) can be found in Crystallography Open Database with ID: 2105418.}] behm1982pentadecacopper}.
Two \textit{bc} and \textit{ac} projections of the triclinic unit cell of \cubo{} are shown in Fig.~\ref{fig:structure}
using the \textsc{vesta} software~\cite{momma2011vesta}.
For a more convenient description of our experimental data, we introduce an additional set of axes, namely, $a$*, $b$*, and $c$*, which are normals to the $bc$-, $ac$-, and $ab$-planes, respectively.
It should be noted that $a$* is the normal to the crystal cleavage planes.
All the atoms build a pseudo-tetragonal nearly-planar layer structure.
All B$^{3+}$ ions have trigonal nearly-planar coordination by O$^{2-}$ ions within the same layer forming [BO$_{3}$] groups.
A part of these triangles are interconnected through the common O$^{2-}$ ions forming [B$_{2}$O$_{5}$] groups.
The Cu$^{2+}$ ions within each layer have nearly planar quadratic coordination by four O$^{2-}$ ions.
Some of the Cu$^{2+}$ ions in the layer are additionally coordinated by one or two O$^{2-}$ ions in the neighboring layers.

Examination of the relevant Cu--O bond lengths (not longer than 3~\AA) within the 16 different positions of Cu$^{2+}$ ions allows us to distinguish three different types of positions, which are shown in Fig.~\ref{fig:structure} by different colors.
These types of copper positions, according to Ref.~\cite{behm1982pentadecacopper},  correspond to different Cu--O polyhedra, namely, i)~a planar square (coordination number~4); ii)~a distorted square pyramid~(4\,+\,1); and iii)~a strongly distorted octahedral coordination~(4\,+\,2).
We will analyze these positions in more detail in Section~\ref{sec:optical_properties} where we discuss the effective coordination numbers and electronic structure of the 3\textit{d}$^9$ states of Cu$^{2+}$ ions in the crystal field.
We remind that the~i) and iii) types of the Cu--O polyhedra are structurally close to two basic groups of Cu$^{2+}$ ions in \cume{}~\cite{martinez1971crystal}.
It is also worth noting that in \cubo{} different kinds of interconnection between groups are realized, for example, vertex-connected [CuO$_4$]--[CuO$_4$], [CuO$_4$]--[CuO$_6$], [CuO$_5$]--[CuO$_6$], and [CuO$_4$]--[BO$_3$] groups, edge-connected [CuO$_6$]--[CuO$_4$], and [CuO$_6$]--[BO$_3$] groups, and others.

\subsection{Magnetic properties}
According to the structural data, the average Cu--O bond length in \cubo{} within the layers is~1.95\,\AA, whereas the bond length between the layers is~2.90\,\AA.
This large difference defines a two-dimensional character of the crystal structure and, consequently, implies a two-dimensional character of the magnetic structure~\cite{petrakovskii1999synthesis}.
Measurements of the magnetic susceptibility~\cite{petrakovskii1999synthesis, petrakovskiui2007magnetic}, specific heat~\cite{petrakovskiui2007magnetic,kudo2001antiferromagnetic}, and inelastic neutron scattering~\cite{petrakovskiui2007magnetic} revealed the formation of a gap of ~$\Delta\sim3.4$\,meV in the spectra of magnetic excitations below the temperature of a magnetic phase transition at \textit{T}$_{N}$\,=\,10\,K, which points to a singlet ground state.
However, magnetic susceptibility measurements along the $a$* axis~\cite{kudo2001antiferromagnetic}, observation of a spin-flop transition at \textit{H}$_{c}$\,=\,9.5\,T~\cite{sakurai2002antiferromagnetic}, and NMR measurements on $^{11}$B nuclei~\cite{sakurai2002antiferromagnetic} point to a 3D antiferromagnetic~(AFM) ordering.
The authors of Ref.~\cite{sakurai2002antiferromagnetic} argued that the ground magnetic state is an AF state with a spin-amplitude modulation similar to a spin-density wave.
A complex magnetic ordering was suggested in~\cite{petrakovskiui2007magnetic} as being a combination of a 3D-AFM ordering and a spin-singlet state.

\section{EXPERIMENTAL DETAILS}\label{sec:exp_details}
In our experiments we used single crystals of \cubo{} grown by L.\,N.~Bezmaternych (Institute of Physics, Krasnoyarsk, Russia) using a spontaneous crystallization method~\cite{petrakovskii1999synthesis}.
Single crystals as large as $\sim$1\,cm$^{3}$ were obtained.
The lattice parameters of the grown crystals were measured with the use of x rays and were in accordance with the literature data for the triclinic structure. 
A perfect cleavage along the $bc$~planes as well as the presence of well-defined $\{001\}$, $\{011\}$, $\{012\}$, $\{031\}$ and other forms could be also identified on the single crystals.
In thin layers, the studied crystals have a deep-green color and noticeable pleochroism (see inset to Fig.~\ref{fig:Abs_inset}), while orthorhombic polymorphs were reported to have a blue color~\cite{zhang2017synthesis}. 
All these observations allowed us to unambiguously assign the grown single crystals to the \textit{P$\overline{1}$} triclinic structure~\cite{behm1982pentadecacopper}, but not to other polymorphs mentioned above.

The investigated crystals of \cubo{} possessed a flaky structure most possibly related to the layered crystalline structure of the compound and weak interactions between the individual structural elements. Such crystals are difficult to analyze by using conventional infrared (IR) spectroscopy because of scattering of the incident light on their rough surfaces resulting in poor quality of the spectra. A much better method for examining such crystals is the so-called attenuated total reflection~(ATR) spectroscopy ~\cite{Harrick1960ATR,Fahrenfort1961ATR}, when almost complete reflection of the light beam from the sample surface is observed. In fact, the radiation slightly penetrates 
inside the sample where it is partially absorbed. Depending on the angle of incidence, the radiation beam can be reflected from the sample surface one or several times. As a result, a signal resembling that in the absorption spectrum is recorded, and the frequencies of the absorbed radiation coincide with the frequencies obtained in the IR transmission spectroscopy. In our experiment, a special Single Reflection ATR Microsampler MVP-Pro with a diamond ATR crystal was used for registering the ATR spectra.

Infrared transmission and ATR spectra were registered in a broad spectral range using a high-resolution Fourier-transform IR spectrometer Bruker IFS 125HR.
Helium-cooled bolometer for the far IR region (10--500\,cm$^{-1}$), and a liquid-nitrogen cooled MCT detector for the middle IR region (400--3000\,cm$^{-1}$) were used.
Samples were cooled in a closed-cycle cryostat Cryomech~ST403. 

Raman spectra were measured in the range ~25--2000\,cm$^{-1}$ with the use of a~T64000~(Jobin-Yvon) spectrometer equipped with a cooled charge-couple device~(CCD) camera.
The line $\lambda$\,=\,532\,nm (2.33\,eV) of a Nd:YAG laser (Laser Quantum) was used as the excitation source with the incident laser power limited to 3\,mW in order to avoid the sample overheating.
All measurements were done in the back-scattering geometry with the use of a $50\times$ objective.
Cooling and stabilization of the sample's temperature between 10 and 300\,K were done with the use of a closed-cycle helium cryostat.
Multiple polarization settings were applied to register the scattering spectra.

For measuring absorption and reflection spectra in the near infrared, visible, and ultraviolet regions, a Bruker IFS\,125HS and a Shimadzu UV-3600\,Plus spectrophotometers were used.

\section{LATTICE DYNAMICS}\label{sec:lattice_dynamics}
\subsection{Symmetry analysis}
The primitive unit cell of triclinic \cubo{} with~\textit{Z}\,=\,10 contains 110~atoms which results in total 330~phonon modes.
The group-theoretical analysis gives the following set of the lattice modes in the center of the Brillouin zone:

\begin{equation}
\label{eq:modes}
330{\Gamma}=162A_g(xx,yy,zz,xy,xz,yz)+168A_u(x,y,z)
\end{equation}

Among them, there are 3$A_u$ acoustic modes; 165$A_u$ odd optical modes are IR-active, and 162$A_g$ even modes are Raman-active.
Low crystal symmetry of \cubo{} allows one to observe Raman-active modes in all diagonal and off-diagonal polarization settings.
The main types of lattice vibrations in the high-symmetry [BO$_{3}$], [CuO$_{4}$], [CuO$_{5}$], and [CuO$_{6}$] groups can be found in~\cite{nakamoto}.
Because of a low symmetry of the crystal as a whole and low site symmetries, all the phonons are nondegenerate.

\subsection{Infrared spectroscopy}
\begin{figure}
\centering
\includegraphics[width=\columnwidth]{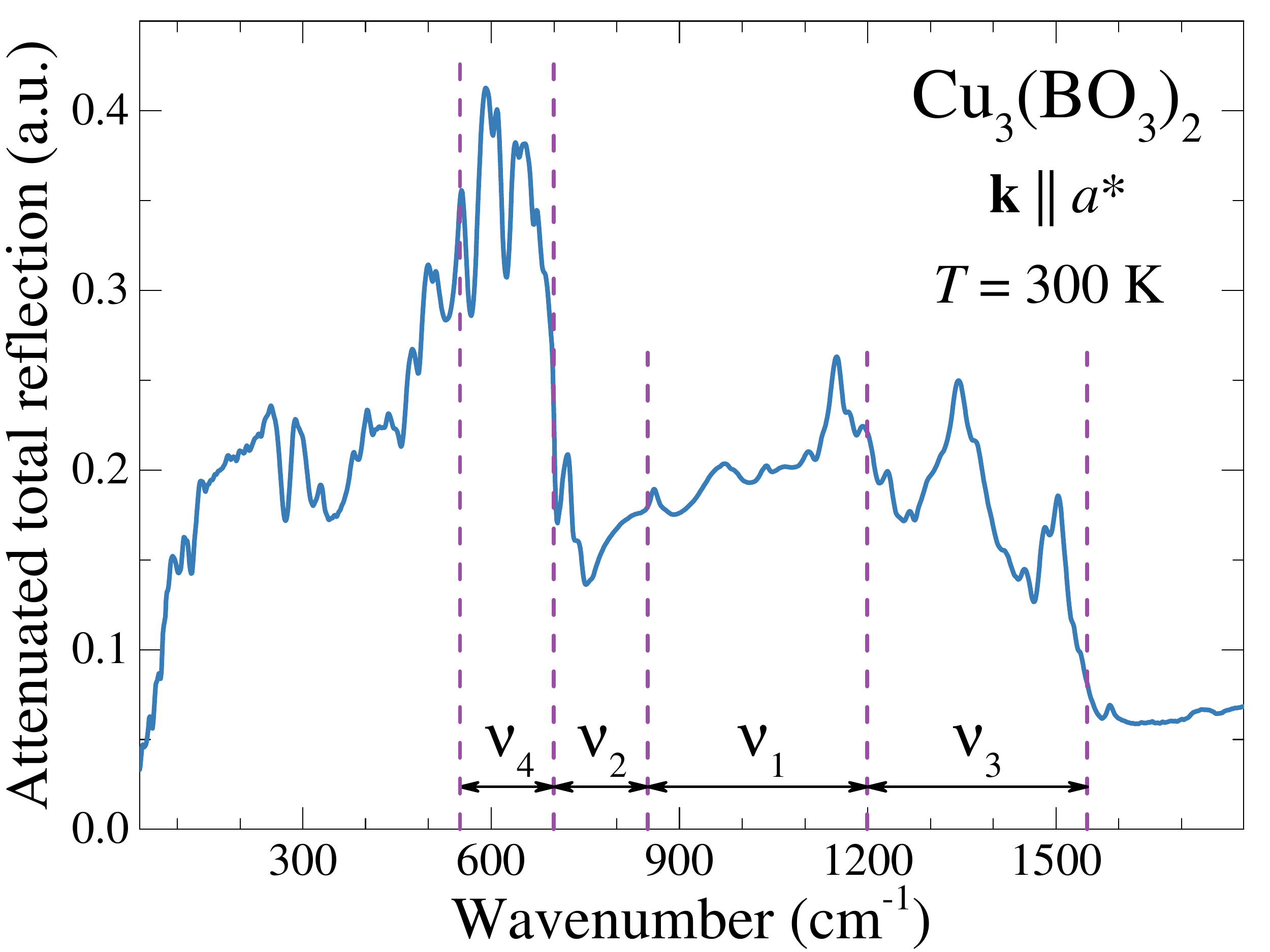}
\caption{\label{fig:ATR}
(Color Online) Attenuated total reflection (ATR) spectrum of \cubo{} in the range 40-–1800\,cm$^{-1}$ at room temperature.
Frequency regions corresponding to internal vibrations of the [BO$_{3}$] groups are indicated.
}
\end{figure}

Fig.~\ref{fig:ATR} shows the ATR spectrum from a cleavage plate of \cubo{} measured at room temperature.
The spectrum demonstrates a very rich phonon structure due to the presence of a large number of atoms in the primitive cell.
One can distinguish several frequency regions related to the quasi-molecular internal vibrations ${\nu}_4$, ${\nu}_2$, ${\nu}_1$, and ${\nu}_3$ of triangular [BO$_{3}$]$^{3-}$ groups~\cite{fausti2006raman, nakamoto}, namely, 550--700\,cm$^{-1}$, 700--850\,cm$^{-1}$, around 1000\,cm$^{-1}$, and 1200--1550\,cm$^{-1}$, respectively.
Vibrations at lower frequencies are connected with translational and librational movements of [BO$_{3}$]$^{3-}$ groups (molecular weight \textit{M}\,=\,59) and movements of copper atoms (\textit{M}\,=\,63).
The observed vibrational spectrum of \cubo{} occupies the frequency range 100--1550\,cm$^{-1}$, which is somewhat broader that the range 100--1200\,cm$^{-1}$ previously observed in \cume{}~\cite{pisarev2013lattice}, the crystal with a similar chemical composition. 
The strongest bonds in both structures are B--O bonds but whereas in \cubo{} these bonds are within BO$_{3}$ planar triangles, in \cume{} they form BO$_{4}$ tetrahedrons.
The frequencies of eigenvibrations of a free BO$_{3}$ molecule are higher than those of a BO$_{4}$ tetrahedral molecule~\cite{nakamoto}, and this can explain why higher frequencies are observed in \cubo{}.
It is worth noting, however, that in the compounds \textit{R}Fe$_{3}$(BO$_{3}$)$_{4}$ (\textit{R} is a rare earth ion), which also contain planar triangular BO$_{3}$ groups, the highest observed vibrational frequency was of about 1440\,cm$^{-1}$.
Even higher frequencies of about 1550\,cm$^{-1}$ in \cubo{} can be tentatively assigned to internal vibrations of tightly bound B$_{2}$O$_{5}$ units. 

\begin{figure}
\centering
\includegraphics[width=\columnwidth]{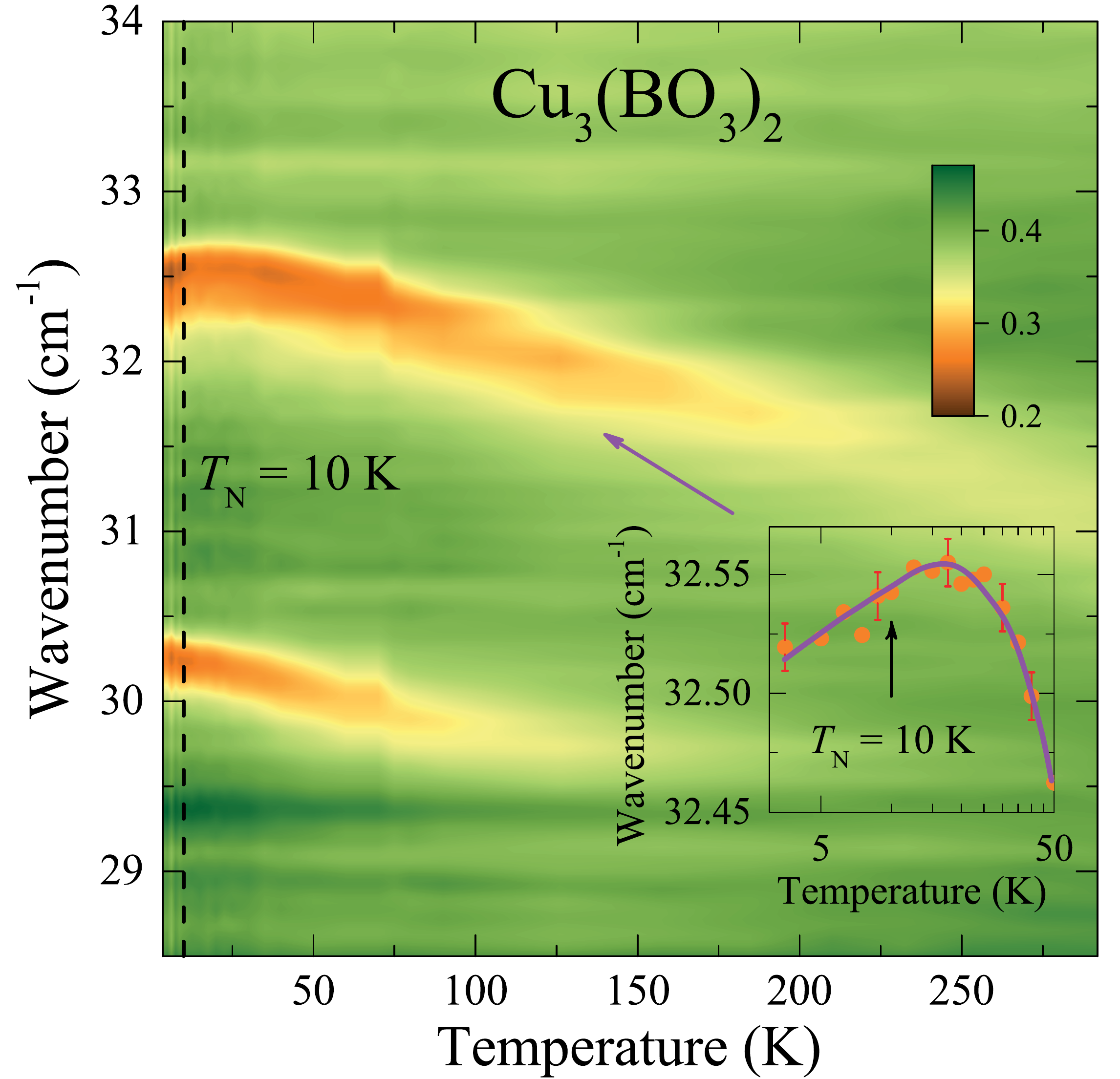}
\caption{\label{fig:two_modes}
(Color Online) Low-frequency transmission spectrum of \cubo{} as a function of temperature.
The values of transmittance are indicated by the color scale in the right upper corner.
Inset shows the temperature dependence of the frequency of the ${\nu}_2$\,=\,32.6\,cm$^{-1}$ mode.
}
\end{figure}

\begin{figure}
\centering
\includegraphics[width=\columnwidth]{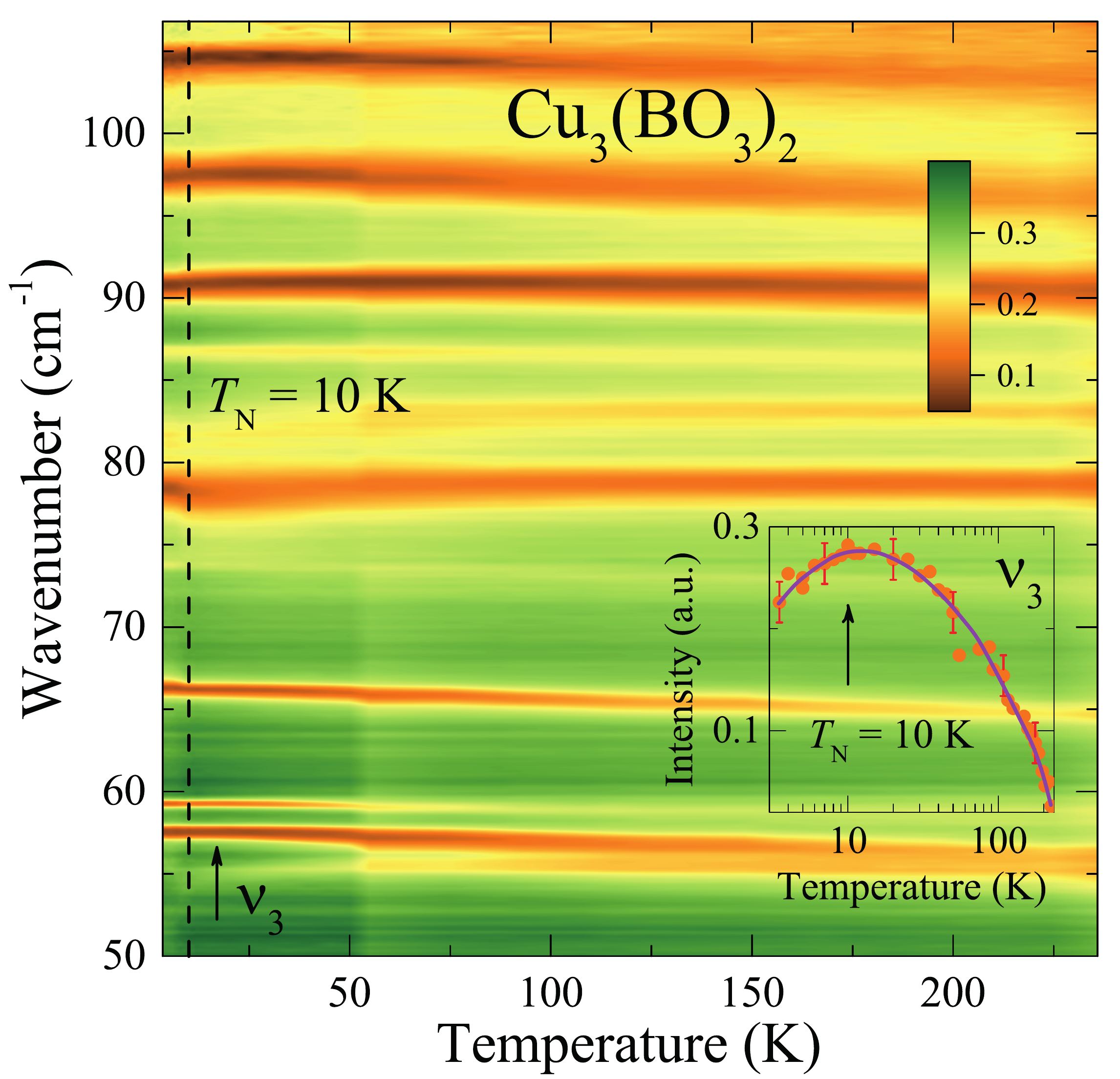}
\caption{\label{fig:phonon_map}
(Color Online) Transmission spectrum of \cubo{} in the spectral region~50-–107\,cm$^{-1}$ as a function of temperature.
The values of transmittance are indicated by the color scale in the right upper corner.
Inset shows the intensity of the ${\nu}_3$\,=\,57.5\,cm$^{-1}$ line as a function of temperature.
}
\end{figure}

The main difference between the infrared spectra of \cubo{} and \cume{} is that for the former one the transmission spectrum contains a large number of lines with frequencies below~100\,cm$^{-1}$. There are two phonon lines with frequencies around~30\,cm$^{-1}$ (see~Fig.~\ref{fig:two_modes}), and about ten lines in the range 55--97\,cm$^{-1}$, (see~Fig.~\ref{fig:phonon_map}).
Low-frequency vibrations are characteristic for layered compounds (to which \cubo{} belongs), because of weak couplings between the layers.
The number of inter-layer optical vibrations (induced by acoustic vibrations of layers) is \textit{N}\,=\,3(\textit{N}$_L$-1)~\cite{kitaev1994quasi}, where \textit{N}$_L$ is the number of layers in the primitive unit cell.
According to the structural data~\cite{behm1982pentadecacopper}, \textit{N}$_L$\,=\,1 for \cubo{}, and, thus, \textit{N}\,=\,0.
However, we may assume that the crystal structure of \cubo{} is even more complex and a superstructure along the \textit{a} axis exists resulting in a corresponding folding of the Brillouin zone.
Acoustic phonons from the zone boundary are thus transferred to the zone center where they become optically active.
The presence of such extra folded modes was observed in the IR spectra of the copper oxide CuO~\cite{kuz2001infrared}.
As the most probable mechanism for the formation of a superstructure in CuO, the authors of~\cite{kuz2001infrared} considered the formation of polar Jahn\,--\,Teller centers [CuO$_{4}$]$^{5-}$ and [CuO$_{4}$]$^{7-}$, according to the model developed earlier for explaining some experimental results on CuO and high-\textit{T}$_c$ cuprates~\cite{moskvin1993pseudo, moskvin1994characteristics}.
One cannot exclude such a mechanism in the case of \cubo{}, but, obviously, the problem requires a further study.

Assumption about the superstructure in \cubo{} allows us to explain in a natural way low-frequency ``extra'' modes in the IR spectra.
Two lowest modes at about~30\,cm$^{-1}$ originate from relative shifts of the two neighboring layers in the \textit{bc}~plane.
The third inter-layer mode corresponding to the relative movement of two layers along the $a$* direction cannot be excited by the light polarized in the \textit{bc}~plane.
Modes with frequencies in the range~55--97\,cm$^{-1}$, most probably, correspond to intra-layer vibrations at specific points of the Brillouin zone, activated by the zone folding.
Extremely small width (of about~0.3\,cm$^{-1}$) of these low-frequency modes and their low intensity (they are not observed in reflection but only in transmission) supports their interpretation as folded modes~\cite{popova1997appearance}.

We consider now the temperature behavior of the observed infrared-active vibrational modes.
Two modes shown in Fig.~\ref{fig:two_modes} harden and narrow with decreasing the temperature.
At \textit{T}\,=\,3.5\,K, ${\nu}_1$\,=\,30.2\,cm$^{-1}$, $\Delta{\nu}_1$\,=\,0.3\,cm$^{-1}$, and ${\nu}_2$\,=\,32.6\,cm$^{-1}$, $\Delta{\nu}_2$\,=\,0.5\,cm$^{-1}$.
Such behavior is typical for the lattice phonons.
However, in the low-temperature region different behavior is observed for the 32.6\,cm$^{-1}$ line, namely, in the vicinity of the magnetic transition at \textit{T}$_{N}$\,=\,10\,K the trend for its frequency change becomes negative, as it is shown in Fig.~\ref{fig:two_modes}, Inset.
Such behavior evidences pronounced magnetoelastic interaction, which leads to a softening of the elastic constants for this particular mode and a softening of the relevant frequency.
The 2D magnetic correlations take place at temperatures well above \textit{T}$_{N}$~\cite{de1974experiments} and this explains why the frequency changes caused by magnetoelastic interactions are noticable above \textit{T}$_{N}$.
Superposition of the two contributions with opposite temperature trends results in the dependence with a maximum at 13\,K.

Similar behavior is observed for some phonons at higher frequencies.
However, several phonons, for example at ${\nu}$\,=\,80\,cm$^{-1}$ in~Fig.~\ref{fig:phonon_map}, demonstrate a slight increase of the frequency in the vicinity of the magnetic phase transition. 
The matter is that the magnetoelastic interaction may result in both softening and hardening of the elastic constants for a particular mode~\cite{kuz2001infrared}.
The magnetoelastic interaction also introduces an additional contribution to anharmonic interactions in the crystal~\cite{klimin2016infrared}, which leads, in particular, to a redistribution of intensities between the modes.
As an example, Inset in Fig.~\ref{fig:phonon_map} shows the temperature dependence of the intensity of the phonon at ${\nu}_3$\,=\,57.5\,cm$^{-1}$, demonstrating a peculiarity at 13\,K.
It is important to note that no new phonon modes were found below \textit{T}$_{N}$, and this observation excludes the spin\,--\,Peierls character of the magnetic transition. 

Table~\ref{tab:IR_frequencies} presents the frequencies of all IR-active modes observed in the IR transmission and ATR spectra.

\begin{table}
\caption{\label{tab:IR_frequencies} Frequencies (cm$^{-1}$) of the IR-active $A_u$ modes observed on the IR transmission and ATR spectra}
\begin{ruledtabular}
\begin{tabular}{cccccccccccccccccccccccccccc}
\multicolumn{28}{c}{IR transmission spectra at $T$\,=\,3.5\,K ($T$\,=\,235\,K)}\\\hline
\multicolumn{7}{c}{30.2 (29.2)}&\multicolumn{7}{c}{59.3 (58.9)}&\multicolumn{7}{c}{78.5 (78.7)}&\multicolumn{7}{c}{90.8 (90.5)}\\
\multicolumn{7}{c}{32.6 (31.3)}&\multicolumn{7}{c}{66.2 (64.7)}&\multicolumn{7}{c}{81.0 (83.0)}&\multicolumn{7}{c}{97.4 (96.3)}\\
\multicolumn{7}{c}{55.3 (55.1)}&\multicolumn{7}{c}{72.2 (72.4)}&\multicolumn{7}{c}{81.7 (-)}&\multicolumn{7}{c}{104.6 (103.3)}\\
\multicolumn{7}{c}{57.5 (55.9)}&\multicolumn{7}{c}{73.8 (72.8)}&\multicolumn{7}{c}{86.8 (86.3)}\\\hline
\multicolumn{28}{c}{ATR spectrum at $T$\,=\,300\,K}\\\hline
\multicolumn{4}{c}{111.2}&\multicolumn{4}{c}{224.4}&\multicolumn{4}{c}{328.3}&\multicolumn{4}{c}{511.9}&\multicolumn{4}{c}{685.7}&\multicolumn{4}{c}{1150.6}&\multicolumn{4}{c}{1420.1}\\
\multicolumn{4}{c}{116.3}&\multicolumn{4}{c}{230.6}&\multicolumn{4}{c}{381.3}&\multicolumn{4}{c}{552.7}&\multicolumn{4}{c}{722.3}&\multicolumn{4}{c}{1170.1}&\multicolumn{4}{c}{1450.7}\\
\multicolumn{4}{c}{138.6}&\multicolumn{4}{c}{240.7}&\multicolumn{4}{c}{403.0}&\multicolumn{4}{c}{591.6}&\multicolumn{4}{c}{738.9}&\multicolumn{4}{c}{1195.7}&\multicolumn{4}{c}{1481.7}\\
\multicolumn{4}{c}{181.2}&\multicolumn{4}{c}{249.3}&\multicolumn{4}{c}{436.9}&\multicolumn{4}{c}{609.3}&\multicolumn{4}{c}{860.0}&\multicolumn{4}{c}{1232.2}&\multicolumn{4}{c}{1503.7}\\
\multicolumn{4}{c}{190.9}&\multicolumn{4}{c}{255.8}&\multicolumn{4}{c}{448.4}&\multicolumn{4}{c}{638.5}&\multicolumn{4}{c}{972.1}&\multicolumn{4}{c}{1267.2}&\multicolumn{4}{c}{1527.5}\\
\multicolumn{4}{c}{200.3}&\multicolumn{4}{c}{288.3}&\multicolumn{4}{c}{474.1}&\multicolumn{4}{c}{653.3}&\multicolumn{4}{c}{1038.1}&\multicolumn{4}{c}{1344.7}&\multicolumn{4}{c}{1539.1}\\
\multicolumn{4}{c}{210.7}&\multicolumn{4}{c}{299.1}&\multicolumn{4}{c}{499.0}&\multicolumn{4}{c}{672.8}&\multicolumn{4}{c}{1105.9}&\multicolumn{4}{c}{1373.5}&\multicolumn{4}{c}{1585.6}\\
\end{tabular}
\end{ruledtabular}
\end{table}

\subsection{Raman spectroscopy}

\begin{figure*}
\includegraphics[width=175mm]{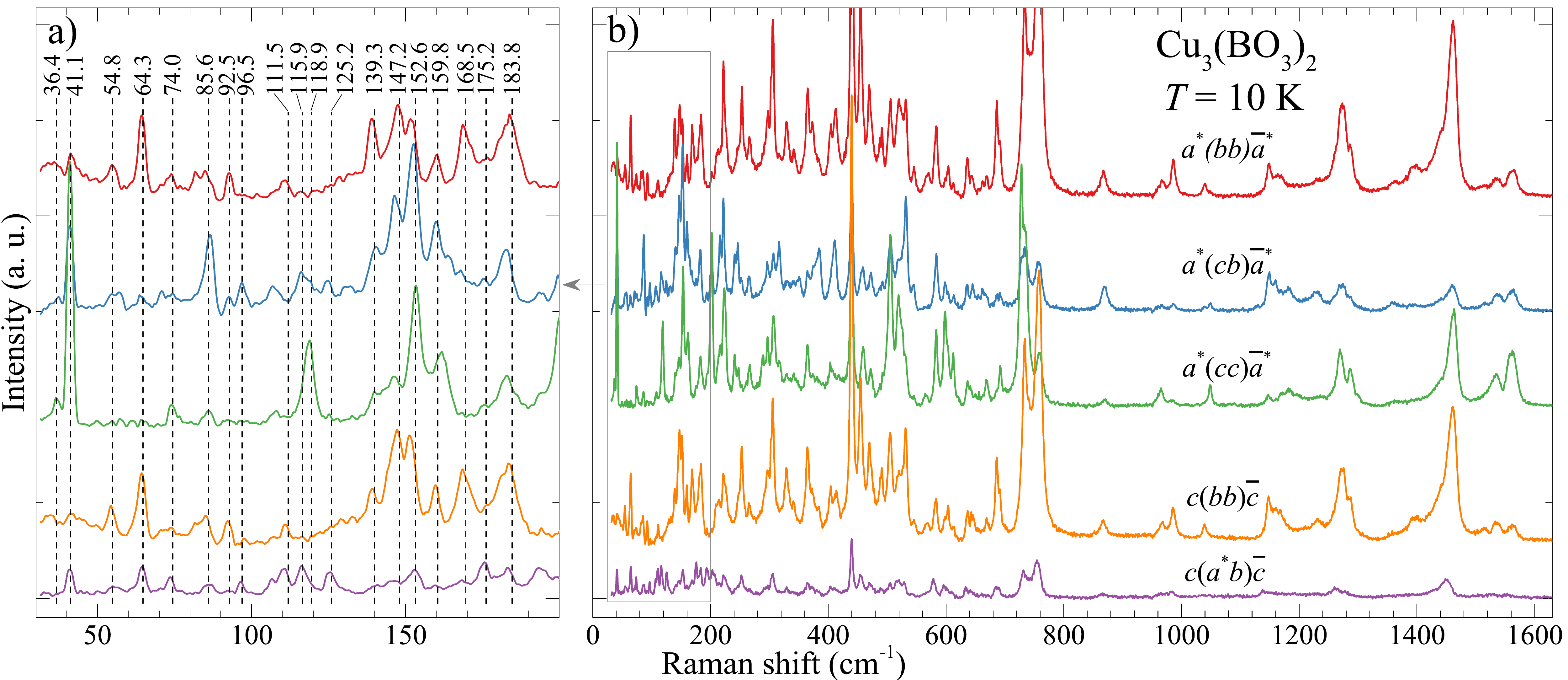}
\caption{\label{fig:Raman}
(Color Online) Raman scattering spectra of \cubo{} for five different polarization settings at \textit{T}\,=\,10\,K.
a)~Low energy  part of the spectrum below $<$\,200\,cm$^{-1}$. Numbers represent Raman shift of several selected intense modes.
b)~Spectra in the whole energy range of~20--1630\,cm$^{-1}$ vertically shifted for clarity.
}
\end{figure*}

Raman studies of low-frequency excitations are always quite a challenging task because of strict requirements to stray-light rejection.
The use of a triple monochromator in dispersion subtraction mode allowed us to register the Raman scattering spectra of \cubo{} beginning from 25~cm$^{-1}$.
They are shown for different polarization settings at $T$\,=\,10\,K in Fig.~\ref{fig:Raman}.
The highest density of the Raman lines is observed in the range 30--800\,cm$^{-1}$, and several more lines are detected at higher energies, with the highest frequency at~1566\,cm$^{-1}$.
Parameters of the all observed phonons were extracted through fitting spectra to Voigt profiles with the use of \textsc{fityk} code \cite{Wojdyr2010fitik} and summarized in Table~\ref{tab:raman_frequencies}.
It is interesting to compare the observed spectra with those of the chemically-related compound \cume{} where the lattice excitation spectra lie in the range  110--1230\,cm$^{-1}$.
An important difference between \cume{} and \cubo{}, besides obvious differences in lattice symmetry, is the coordination of boron, namely, in the former there are [BO$_4$] tetrahedral groups while in the latter one finds either isolated [BO$_3$] groups or [B$_2$O$_5$] groups with shorter B--O bonds.
Thus, it is quite reasonable to expect higher-frequency modes in the lattice excitation spectra of \cubo{}.
Some other examples of Raman spectra of characteristic [BO$_3$] and [B$_2$O$_5$] groups in other borate crystals and melts with phonon modes up to 1500\,cm$^{-1}$ can be found in~\cite{bues1966strukturen}.
Obviously, this assignment of the high-frequency modes to [BO$_3$] or [B$_2$O$_5$] groups does not exclude a possibility that some of them could be related to two-phonon excitations.
Some similarities between \cubo{} and \cume{} spectra can be noticed, e.g., characteristic and intense ``triplet'' at 727-733-757\,cm$^{-1}$ similar to a triplet at 695-704-725\,cm$^{-1}$ in \cume{} but shifted to a lower frequency by~30\,cm$^{-1}$.

It is also worth noting, that, in contrast to \cume{} with $T{_N}$=21\,K in which a strong two-magnon scattering centered at about 80\,cm$^{-1}$ was observed~\cite{ivanov2013ramancub2o4}, no magnetic scattering was detected in \cubo{} at temperatures down to 10\,K.
First of all, that might be due to a much lower $T{_N}$\,=\,10\,K in this compound.
Secondly, the magnetic structure of \cubo{} is completely different, much more complex, and only partially ordered.
Thus, we expect that a two-magnon scattering in \cubo{} should be positioned at much lower energy and smeared over a broader frequency range due to a greater number of spin-wave branches and magnon-magnon interaction.
We suppose that all these factors make two-magnon scattering in \cubo{} indistinguishable from the background. 



\begin{table}
\caption{\label{tab:raman_frequencies} Frequencies (cm$^{-1}$) of the experimentally observed Raman active $A_g$ modes extracted from the spectra with different polarizations at $T$\,=\,10\,K.}
\begin{ruledtabular}
\begin{tabular}{ccccccc}
36.4&118.9&215.3&350.6&525.1&691.0&1228.5\\
41.1&125.2&222.6&364.4&531.5&733.1&1269.7\\
54.2&139.3&240.8&370.2&546.0&743.4&1272.0\\
64.3&145.8&246.0&376.3&566.9&757.6&1287.8\\
71.5&147.2&250.5&384.7&580.1&850.5&1362.8\\
74.0&152.6&265.7&403.2&601.2&869.1&1369.3\\
81.9&159.8&288.9&412.3&612.1&966.8&1396.8\\
85.6&168.5&296.2&439.6&636.1&985.2&1399.6\\
92.5&175.2&305.5&456.5&644.4&1043.3&1462.4\\
96.5&183.8&316.3&470.5&650.8&1147.5&1531.6\\
106.5&193.5&329.2&491.4&661.9&1158.7&1536.1\\
111.5&203.0&336.1&504.5&668.6&1166.1&1539.6\\
115.9&211.7&341.4&518.9&685.8&1181.2&1565.9\\
\end{tabular}
\end{ruledtabular}
\end{table}

All nine Raman-tensor elements are non-zero for triclinic crystals, and therefore all phonons should be observed at any polarization; nonequivalence of these elements leads to a variation of intensities in the spectra in different polarization settings.
For example, one of the lowest-frequency phonon at 41.1\,cm$^{-1}$ is the most pronounced in the $a$*($cc$)$\overline{a}$* polarization.


The rich set of low-frequency phonons in \cubo{} described above deserves a further discussion. 
Typically, low-frequency phonons can be observed in systems containing heavy ions, e.g. ZrW$_2$O$_8$~\cite{ernst1998phonon}, or as soft modes in materials with structural and/or ferroelectric transitions. 
An example of a material containing both heavy ions and soft modes is the lead zirconate PbZrO$_3$ with several modes below 100~cm$^{-1}$~\cite{hlinka2014multiple}.
Another possibility is often found in organic systems with phonon modes characterized by extremely low force constants.
The cystine is a good example in which a bunch of phonon lines below 100~cm$^{-1}$ is observed  corresponding mainly to torsion vibrations of large organic groups~\cite{brandt2008terahertz}. 
This compound is even used as a calibrant for ultra-low frequency Raman scattering measurements.
Another example is Cu$_3$Mo$_2$O$_9$ in which phonons with frequencies as low as 31~cm$^{-1}$ were observed and tentatively assigned to rigid-chain modes when each of the structure-forming chains moves as a rigid unit~\cite{sato2014raman}.
Since \cubo{} is composed of relatively light ions and cannot be considered as a molecular or soft-mode crystal, alternatively to the folding mechanism discussed in the previous subsection, the observed low-frequency modes could be assigned to external vibrations of individual or coupled groups due to the low symmetry and large unit cell.
Evidently, these interesting observations require further experimental and theoretical studies.
It should be noted that computation of lattice dynamics properties of such a low-symmetry compound with a very big unit cell is an extremely challenging task not only for \textit{ab initio} methods but also for (semi-)classical force-field methods due to rich variety of different coordination polyhedra.
That is a challenging task but it could be of a great help in understanding the nature of low-frequency part of the lattice dynamics in \cubo{}.

The above-discussed low-frequency IR and Raman phonons in \cubo{} lie in the terahertz frequency range (1--5\,THz) and thus can provide a playground for either ultrafast optical excitation of such modes by femtosecond laser pulses or for selective THz excitation of particular modes.
Such methods were applied for phonon-assisted phase transitions or even for a modulation of the exchange interaction~\cite{mikhaylovskiy2015ultrafast}.

\section{ELECTRONIC STRUCTURE AND OPTICAL PROPERTIES}\label{sec:optical_properties}
\subsection{Analysis of crystallographic positions of Cu$^{2+}$ ions in \cubo{}}

In \cubo{}, the Cu$^{2+}$ ions with the 3$d^9$ orbitals occupy sixteen nonequivalent crystallographic positions of three types~\cite{behm1982pentadecacopper} as shown in Fig.~\ref{fig:structure}.
Among them, four are planar-square positions with the four nearest oxygens O$^{2-}$; two are distorted square pyramids with (4\,+\,1) oxygens; and ten are strongly distorted octahedral positions with (4\,+\,2) oxygens.
However, such conclusions about the coordination of Cu ions were based on a somewhat arbitrary basis.
While the well-known coordination-number scheme works well in the case of regular polyhedra it becomes not quite applicable in some cases.
In fact, in \cubo{} there is a large number of inequivalent positions with irregular and strongly distorted Cu--O polyhedra and therefore it becomes more preferable to use the so-called \textit{effective coordination number} (ECoN) approach~\cite{giacovazzo2002fundamentals,hoppe1989new}.
This approach is based on adding weighting scheme for selected bonds, where the fractions between 0 and 1 are assigned depending on its length.
For calculating $N_{ECoN}$ one needs to know only bond lengths of the polyhedra under analysis~\cite{giacovazzo2002fundamentals}.

As the first step, the weighted average bond length should be calculated according to the equation
\begin{equation}
\label{eq:average_bond}
l_{av}=\frac{\sum_i l_i \exp \big[1-(l_i/l_{min})^6\big]}{\sum_i \exp \big[1-(l_i/l_{min})^6\big]},
\end{equation}
where $l_{min}$ is the length of the shortest bond in the polyhedron. The subsequent steps are aimed on calculating the relevant weights for each bond and the sum of these weighted bonds gives the $N_{ECoN}$ as
\begin{equation}
\label{eq:econ}
N_{ECoN} = \sum_i \exp \Big[ 1 - \Big( \frac{l_i}{l_{av}} \Big)^6 \Big].
\end{equation}

\begin{figure}
\centering
\includegraphics[width=\columnwidth]{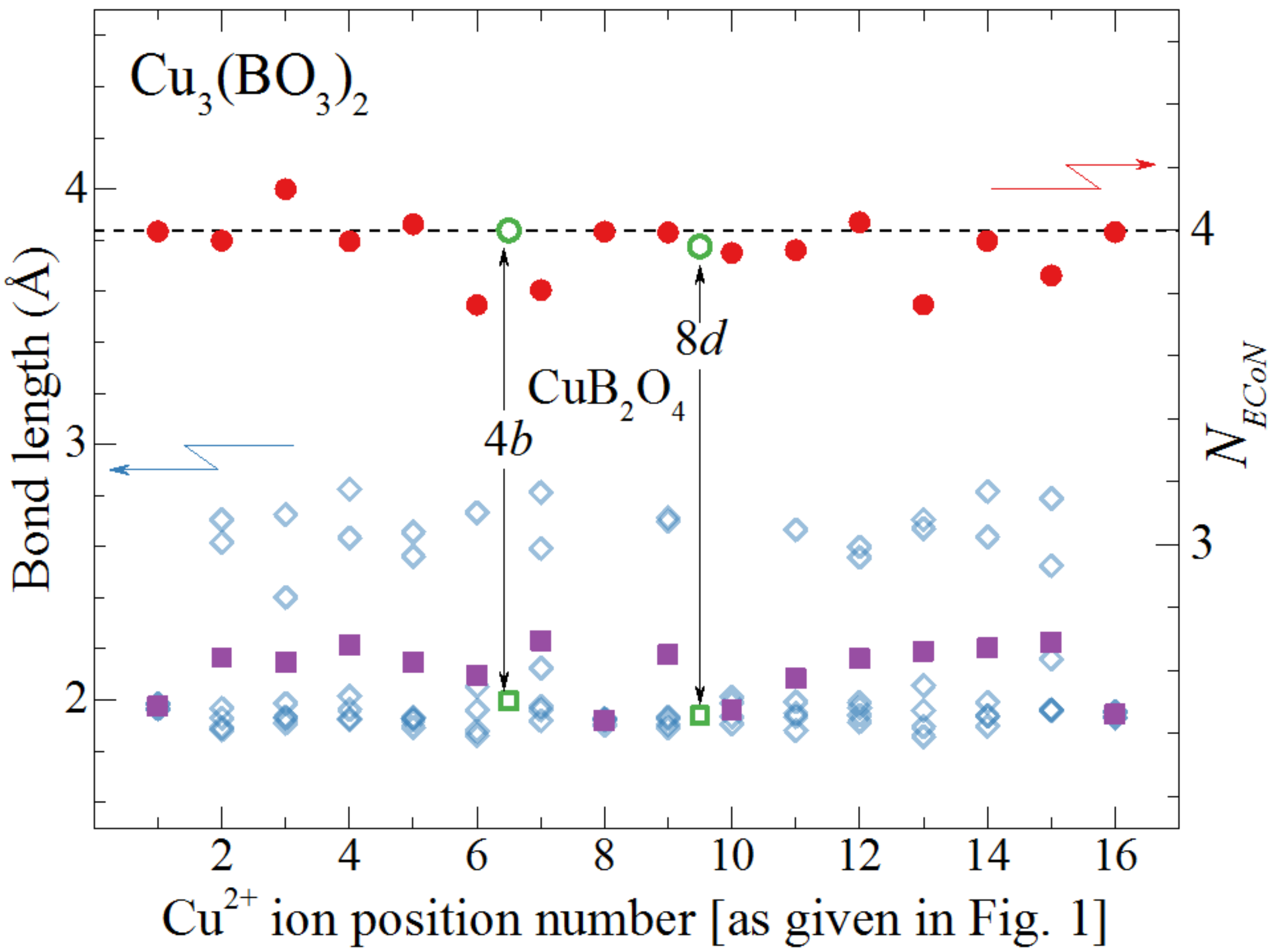}
\caption{\label{fig:ECoN}
Analysis of the sixteen crystallographic positions of Cu$^{2+}$ ions  in \cubo{} (see Fig.~\ref{fig:structure}) for the nearest Cu--O bonds with the length up to 3~\AA ~according to the crystallographic data from~\cite{behm1982pentadecacopper}.
Filled purple squares show averaged bond lengths for all polyhedra.
Empty diamonds represent the length of all bonds.
Red filled circles show calculated values of effective coordination numbers.
Dashed line shows the value of $N_{ECoN}$\,=\,4 for a perfectly square-planar coordination. 
Green symbols show the $N_{ECoN}$ values for two different crystallographic positions, 4$b$ and 8$d$, of \cume{}~\cite{abdullaev1981refined}.
}
\end{figure}

Such analysis was performed for the all sixteen positions of Cu$^{2+}$ ions in \cubo{} presented in Fig.~\ref{fig:structure}.
Calculated values of $N_{ECoN}$~along with average bond lengths $l_{av}$ ~are shown in~Fig.~\ref{fig:ECoN}.
The analysis shows that the average bond lengths for all positions are close to $l_{av}$\,=\,2.116\,\AA~ while the average effective coordination number $N_{avECoN}$\,=\,3.94.
Thus, this analysis allows us to describe all the Cu-containing groups as fourfold-coordinated nearly square-planar groups. We performed such analysis for some other cuprates.
For example, $N_{ECoN}$\,=\,4.0 in the monoclinic CuO~\cite{tunell1935geometrical}. 
Analyzing the structural data of several other cuprates we have found that the value of $N_{ECoN}$\,=\,4.0 and nearby is typical for many other Cu$^{2+}$ complex oxides.
But there are some interesting exceptions.
For copper ions in the above-mentioned orthorhombic \cubo{}~\cite{zhang2017synthesis} which is structurally equivalent to other orthorhombic borates~\cite{pisarev2016lattice}, $N_{ECoN}$ is very close to six for both types (2\textit{b} and 4\textit{d}) of the [Cu--O] groups.
We note that the triclinic \cubo{} where $N_{ECoN}\approx4$ and the orthorhombic phase where $N_{ECoN}\approx6$ have markedly different density, namely \(\rho\)\,=\,4.54  and 5.12\,g/cm$^3$, respectively, while the density of monoclinic phase is 4.434\,g/cm$^3$.
It should be noted that such difference in densities of chemically complex compounds is quite a rare phenomenon.
The difference in the effective coordination of copper ions leads to a markedly more friable structure of the triclinic \cubo{}.
This comparison of the two materials with the same chemical composition but crystallizing in different crystal structures and characterized by different $N_{ECoN}$ values emphasizes once more the remarkable ability of Cu$^{2+}$ ions for adopting different positions in the unit cell, preferably, with a lower local symmetry as a result of the Jahn\,--\,Teller effect.

\subsection {Optical absorption in the region of \textit{d-d} electronic transitions}

\begin{figure}
\centering
\includegraphics[width=\columnwidth]{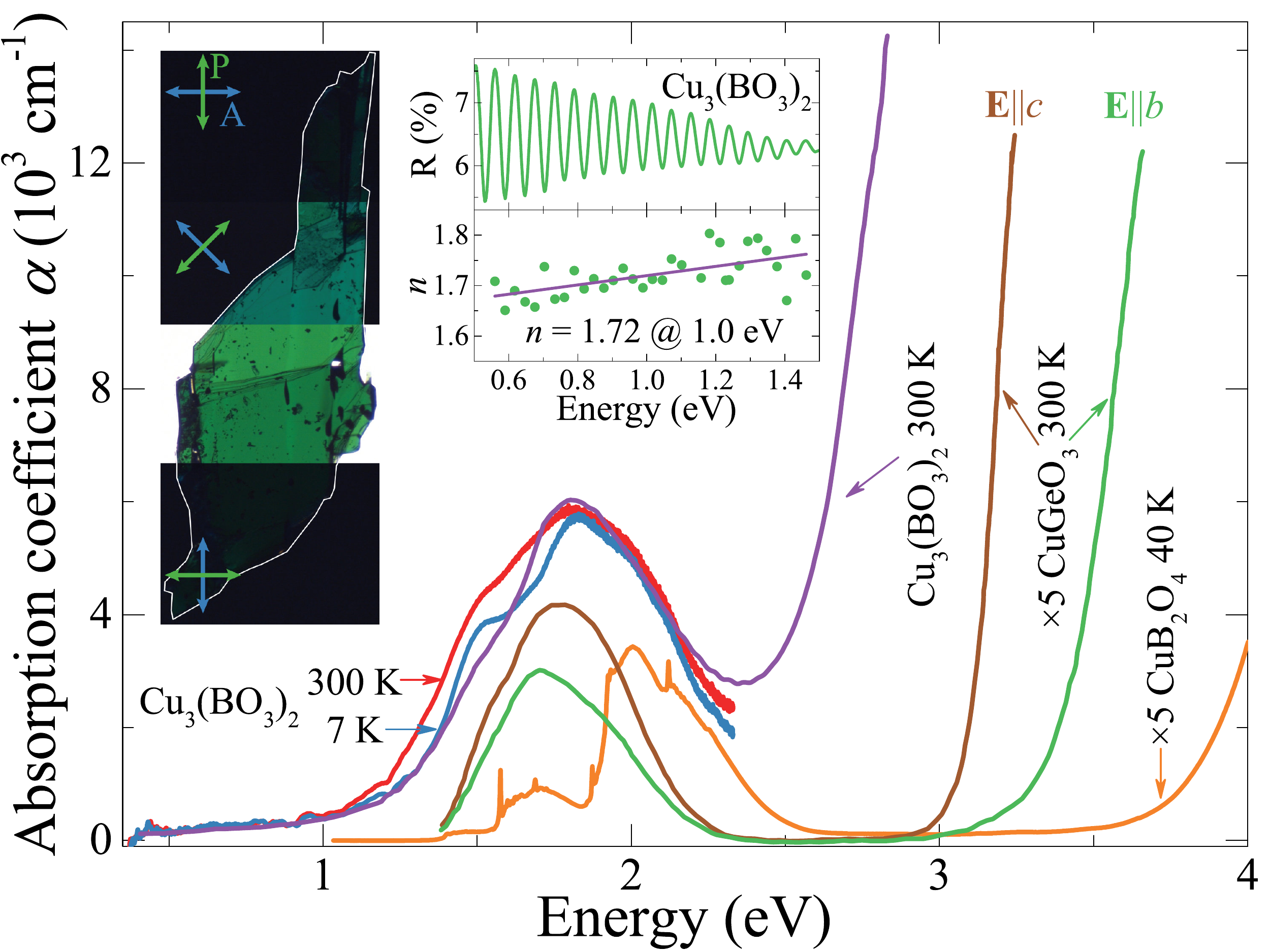}
\caption{\label{fig:Abs_inset}
(Color online) Absorption spectra of \cubo{} at \textit{T}\,=\,7 and 300\,K (\textbf{k}$\parallel$\textit{a}*, unpolarized), of \cume{} at 40\,K (\textbf{k}$\parallel$\textit{c}, unpolarized), and of \cugeo{} at 300\,K for two polarizations~\cite{popova1996optical}.
The photo at the left-hand side shows a thin sample of \cubo{} cleaved perpendicular to the \textit{a}* axis
and placed between the crossed polarizers.
Green and blue arrows represent the angles of a polarizer and analyzer, respectively.
Inset: Fringes in the reflected light due to the interference of the beams reflected from the crystal faces in the transparency window (upper panel) and the refractive index calculated from these fringes (lower panel).
}
\end{figure}

The crystal-field analysis of the 3$d^9$ orbitals (3$d^1$ in the hole representation) for the all types of positions shows that the ground state is \textit{x}$^2$-\textit{y}$^2$.
Obviously, the relative energy positions of the excited states vary from one position to another because they are defined by such factors as i) the symmetry of the position and ii) specific values of the crystal-field parameters.
Electronic \textit{d-d} transitions from the ground state to excited states are forbidden by the parity selection rule, in particular, in the square-planar position possessing the inversion center $\overline{1}$ (position numbers 1 and 16 in Fig.~\ref{fig:structure}).
However, this rule is broken in square pyramids and distorted octahedra, due to admixture of wave functions with different parity (from excited configurations of copper or from charge-transfer bands) to the \textit{d} wave functions by the odd crystal-field parameters.
Another important contribution to the \textit{d-d} absorption intensity may come from vibronic transitions caused by the interaction between \textit{d} electrons of Cu$^{2+}$ and odd-parity vibrations of the nearest surrounding.
This temperature-dependent mechanism dominates in the optical absorption of~\cugeo{}~\cite{bassi1996optical,popova1996optical}.
For \cubo{}, however, the intensity of the optical \textit{d-d} absorption bands (situated in the spectral region 1.0--2.5\,eV) practically does not depend on the temperature (see Fig.~\ref{fig:Abs_inset}), which confirms a dominant role of a mixing between the 3\textit{d} states of Cu$^{2+}$ and the states of a different parity in the Cu$^{2+}$ centers without the inversion symmetry.

Only the $bc$-plane-cleaved samples of \cubo{} were used in absorption and reflection measurements.
Thin samples were obtained due to the perfect cleavage of the single crystals on the planes shown in Fig.~\ref{fig:structure}.
They were of a dark-green color with a noticeable pleochroism in the visible spectral range when they had the thickness \textit{t}~$\leq$~10--12\,$\mu$m (see the photo at the left side of Fig.~\ref{fig:Abs_inset}).
The room-temperature spectra shown in Fig.~\ref{fig:Abs_inset} compare the optical absorption of \cubo{} (\textbf{k}$\parallel$\textit{a}*, unpolarized), \cume{} (\textbf{k}$\parallel$\textit{c}, unpolarized), and \cugeo{} (\textbf{k}$\parallel$\textit{a}, \textbf{E}$\parallel$\textit{b}, and \textbf{E}$\parallel$\textit{c}) in the regions of both intraconfigurational \textit{d-d} transitions and the fundamental absorption edge.

\begin{figure}
\centering
\includegraphics[width=\columnwidth]{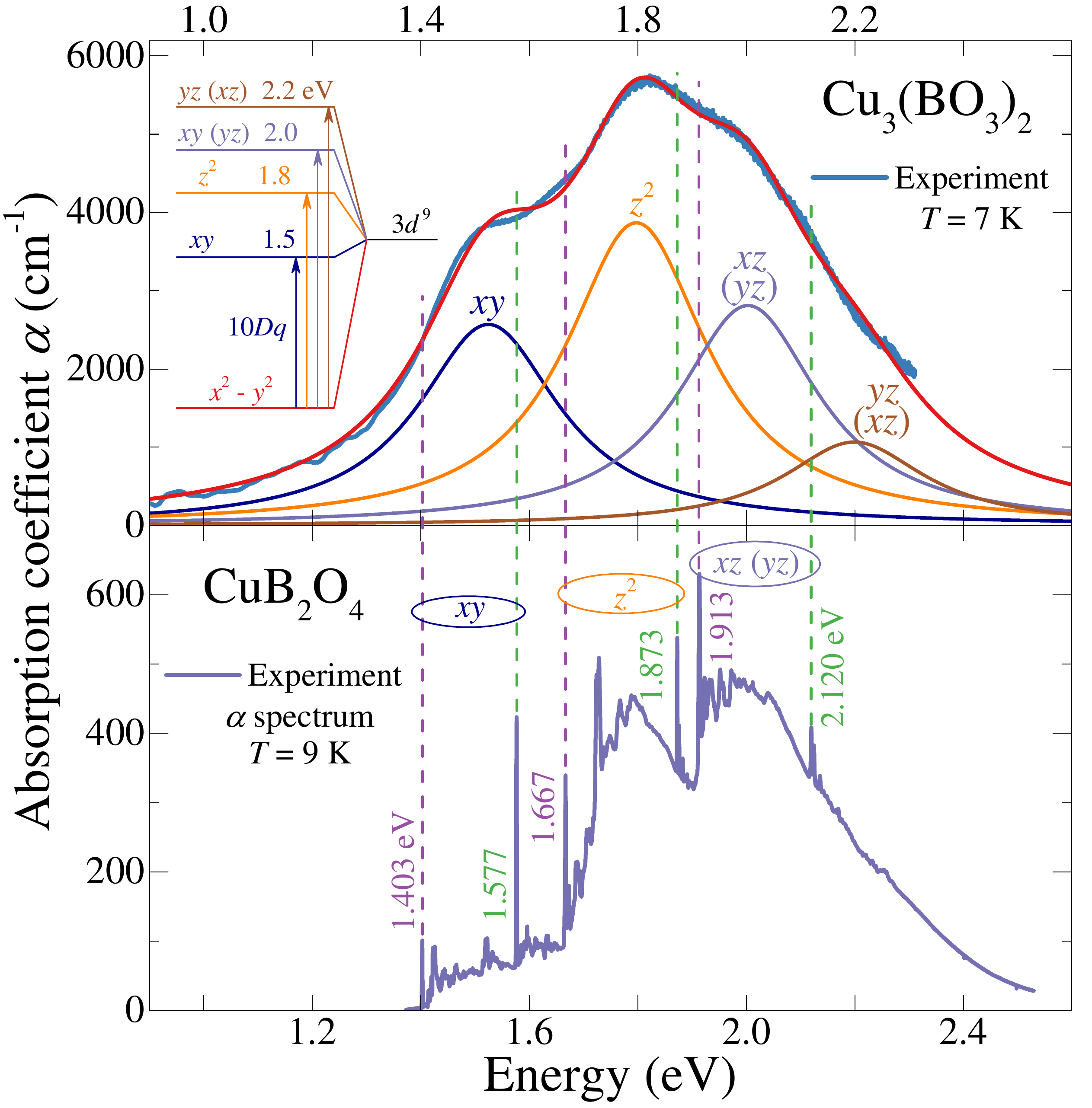}
\caption{\label{fig:Abs}
(Color online) Comparison of electronic absorption spectra of \cubo{} (upper panel) and \cume{} (lower panel) at 7 and 9\,K, respectively.
Inset shows a schematic representation of the nondegenerate energy levels of the Cu$^{2+}$ ions in \cubo{}.
Comparison of the two spectra shows a complete absence of any fine structure in \cubo{}.
On the other hand, the positions of maxima of the absorption bands in \cubo{} (upper panel) lie very closely to the positions of the relevant bands in \cume{} (lower panel).
We note that the \textit{xz} and \textit{yz} states in \cume{} are degenerate because of a high symmetry of the both Cu$^{2+}$ positions whereas in \cubo{} these states are split.
}
\end{figure}

We concentrate first on the spectra of the \textit{d-d} transitions displayed in more detail in Fig.~\ref{fig:Abs}, upper panel.
For \cubo{} (and also \cugeo{}) they remain broad when the temperature decreases but peculiarities of the spectral shape become more pronounced.

Absolute values of the absorption coefficient were calculated from measurements on several samples with the thickness of 5--6\,$\mu$m, and we estimate the accuracy of the measured absorption coefficient as not better than $\sim$\,10\,{\%}.
At the maximum (1.82\,eV), the absorption coefficient reaches the value of $\alpha$\,=\,5800\,cm$^{-1}$.
Strong anisotropy of the optical absorption and index of refraction is expected but this is a subject of future studies.
For comparison, the lower panel in Fig.~\ref{fig:Abs} shows the low-temperature absorption spectrum of \cume~\cite{pisarev2004magnetic}, which radically differs from that of~\cubo{}.
First, characteristic rich fine structure of the electronic spectrum of \cume{} with well-defined ZP lines and sharp phonon sidebands at the both crystallographic positions completely disappears in \cubo{}.
We note that no ZP lines or any fine structure were observed in a structurally more simple \cugeo{}, the spectrum of which~\cite{bassi1996optical,van1998optical,popova1996optical} is shown in Fig.~\ref{fig:Abs_inset}.
Second, the absorption coefficient at the maximum is roughly an order of magnitude larger for \cubo{} than for \cume{} and \cugeo{}.
On the other hand, the value $\alpha$\,=\,5800\,cm$^{-1}$ falls in the range of absorption coefficients found in several other Cu$^{2+}$-based oxide minerals, see, for example, Table~5.20 in Ref.~\cite{burns1993mineralogical}. 

A question arises why there are such drastic differences between the absorption spectra of chemically similar borates \cubo{} and \cume{}.
Several factors can be assumed.
First of all, \cume{} should be considered as a more ionic compound whereas \cubo{} is more covalent, due to relatively larger concentration of Cu--O bonds by a factor of three to one.
We suggest that this factor along with the presence of a larger number of nonequivalent positions lead to a broadening of narrow ZP lines and vibronic transitions in \cubo{} in which only broad overlapping bands are observed.
A low crystal-field symmetry of all Cu$^{2+}$ positions in this compound, on the one hand, leads to an appreciable admixture of excited electronic configurations of the opposite parity to the 3\textit{d}$^9$ configuration, thus enhancing the intensity of the \textit{d-d} transitions.
On the other hand, the low local symmetry completely removes the degeneracy of the 3\textit{d} states of Cu$^{2+}$ ions.
In the upper panel of Fig.~\ref{fig:Abs}, we show results of the fitting of the experimental spectrum by four orbital subbands related to the electronic transitions from the ground \textit{x}$^2$-\textit{y}$^2$ state to the \textit{xy}, \textit{z}$^2$, \textit{xz}\,(\textit{yz}), and \textit{yz}\,(\textit{xz}) excited states (see inset in upper panel).
Broken lines connecting the zero-phonon lines observed in \cume{} and the broad spectrum of \cubo{} show that the maxima of partial bands in this compound are shifted from the relevant zero-phonon lines in \cume{} by the value of 0.08--0.12\,eV lying in the energy range of the whole phonon spectrum.
Position of the maximum of the first partial band defines the cubic crystal-field parameter 10\textit{Dq}\,=\,1.5\,eV averaged over all 16 inequivalent positions of Cu$^{2+}$ ions.
This 10\textit{Dq} value is very close to  the cubic crystal-field parameter in several Cu$^{2+}$-based minerals given in Table~5.20 in Ref.~\cite{burns1993mineralogical}.    
For comparison, the first ZP lines in \cume{} in the 4\textit{b} and 8\textit{d} positions define the genuine cubic parameter 10\textit{Dq}\,=\,1.403\,eV and 1.577\,eV, respectively.

Figure ~\ref{fig:Abs_inset} demonstrates another important difference in the optical properties of \cubo{} and \cume{}.
A significant red shift of the fundamental absorption edge to a value of about 2.8\,eV takes place in \cubo{} in comparison to about 4\,eV in \cume{}.
We can explain such shift by a larger relative concentration of Cu--O bonds (approximate ratio is three to one) in \cubo{}.
In fact, in pure copper oxide CuO, which is usually characterized as a semiconductor with a direct fundamental band gap, the fundamental absorption edge is positioned at about 1.9\,eV~\cite{ray2001preparation,tahir2012electronic,ching1989ground}.
This comparison allows us to make a conclusion that, in Cu$^{2+}$-containing compounds, there is a systematic shift of the absorption edge due to charge-transfer transitions, from the lowest value of about 1.4\,eV to more than 4.0\,eV as the concentration of Cu--O bonds  decreases.
This strong red shift of the absorption edge is due to an increased hybridization between the 3\textit{d} even states of the copper ions and 4\textit{p} odd states of the oxygen ions.
The same hybridization contributes to an enhancement of the \textit{d-d} absorption intensity, which explains in part a strong increase of the electronic absorption up to $\alpha$\,=\,5800\,cm$^{-1}$.
Polarization measurements on the $bc$-plane-cleaved sample depicted at the left side of Fig.~\ref{fig:Abs_inset}, show that extinction angle is close to the angle between the $b$ and $c$ axis, 89$^{\circ}$ and 88.77$^{\circ}$, respectively.

\section{CONCLUSIONS}\label{sec:disscussion}
The present study of the triclinic copper borate \cubo{} has shown that this crystal possesses several unusual properties which, to a large extent, are related to its complex crystallographic layered structure and a large unit cell with~\textit{Z}\,=\,10.
A bright feature of this structure is a perfect cleavage on the $bc$-type crystallographic planes. 
We suppose that a decisive role in forming the structure itself and the properties of the compound belongs to the Jahn\,--\,Teller magnetic  copper Cu$^{2+}$ ions (\textit{S}\,=\,1/2) occupying sixteen strongly distorted nonequivalent positions.
In the original publication~\cite{behm1982pentadecacopper}, these positions were divided into the three main types, on the basis of the number of their nearest neighbors.
However, our analysis of all types of the distorted polyhedra around  Cu$^{2+}$ ions in \cubo{}, on the basis of an effective coordination number (ECoN) approach~\cite{giacovazzo2002fundamentals}, has shown that $N_{ECoN}$ is close to 4 for all the positions, confirming that all of the Cu--O groups can be treated as nearly planar-square [CuO$_4$] groups.
The large difference, more than 10\%, in densities of triclinic, monoclinic, and orthorhombic phases of \cubo{} may point to an important role of Jahn\,--\,Teller distortions in stabilizing each particular phase.

The lattice dynamics of \cubo{} was studied with the use of the infrared absorption and attenuated total reflection, and Raman scattering spectroscopies.
These studies confirmed the presence of a very complex structure of lattice excitations in both infrared and Raman spectra.
In particular, unusual low-frequency phonons were found below~100\,cm$^{-1}$, which was not observed previously in the chemically similar crystal \cume{}~\cite{pisarev2004magnetic}.
At the present state of our research we can only tentatively assign these low-frequency phonons either to interlayer vibrations activated by the presence of the crystal superstructure or to phonons responsible for the movement of large groups of ions.
Perhaps, a more definite explanation of this unusual low-frequency spectrum could be given as a result of theoretical calculations of phonons in this complex and low-symmetry structure with a very large unit cell containing 110~atoms. 
Some of phonons show an anomalous behavior in the vicinity of the magnetic phase transition at \textit{T}$_N$\,=\,10\,K evidencing a pronounced magnetoelastic interaction.
We should emphasize that no new phonons were found below \textit{T}$_N$, which excludes the spin \textit{S}\,=\,1/2 dimerization through a magnetostructural phase transition of the spin\,--\,Peierls type observed, e.g., in \cugeo{}~\cite{kuroe1994raman,damascelli1997infrared,popova1998folded}.

Absorption and reflection measurements in the range 0.35--6.7\,eV allowed us to characterize the electronic structure formed by the 3\textit{d}$^9$ states of Cu$^{2+}$ ions in the low-symmetry crystal fields.
A large number of nonequivalent positions of the Cu$^{2+}$ ions results in a broadening of the range of the crystal-field parameters and a corresponding broadening and averaging of the whole 3\textit{d} absorption spectrum originating from optical transitions in different ion positions.

In a striking contrast to \cume{}, no zero-phonon lines and fine phonon-sideband structure was found in the optical absorption spectra of \cubo{}  characterized by an intense band positioned at about~1.8\,eV, with the maximum absorption coefficient $\alpha$\,=\,5800\,cm$^{-1}$.
Thus, our study confirmed once more that \cume{} remains, to the best of our knowledge, the only known compound in which all zero-phonon lines at the both types of crystallographic positions can be identified, allowing calculations of the genuine crystal field parameters~\cite{pisarev2011electronic}.  
The strong absorption band centered at 1.8\,eV in \cubo{} was assigned to overlapping electronic transitions in all nonequivalent positions from the ground state \textit{x}$^2$-\textit{y}$^2$ of the Cu$^{2+}$ ions to the \textit{xy}, \textit{z}$^2$, \textit{xz}\,(\textit{yz}), \textit{yz}\,(\textit{xz}) excited states.
The first transition defines the cubic crystal-field parameter 10\textit{Dq}\,=\,1.5\,eV (see inset to the upper panel, Fig.~\ref{fig:Abs}).
We suppose that the strong increase of absorption related to these electronic transitions below the fundamental band edge by about an order of magnitude in comparison to \cume{} is due to a larger relative concentration of strongly-covalent Cu--O bonds in \cubo{}, in an approximate ratio three to one.
The stronger covalency also explains a noticeable red shift of the fundamental absorption edge to a value of about $\sim$\,2.8--3.0\,eV in \cubo{}, in comparison with a value of about $\sim$\,4.0\,eV in \cume{}.

\section{ACKNOWLEDGEMENTS}
This work was supported by the Russian Science Foundation under the project No.~16-12-10456. K.N.B. and M.N.P. acknowledge a support of the Russian Academy of Sciences under the Programs for Basic Research ``Topical problems of the low-temperature physics''.

\bibliography{Cu3_BO3_2}

\begin{thebibliography}{68}%
\makeatletter
\providecommand \@ifxundefined [1]{%
 \@ifx{#1\undefined}
}%
\providecommand \@ifnum [1]{%
 \ifnum #1\expandafter \@firstoftwo
 \else \expandafter \@secondoftwo
 \fi
}%
\providecommand \@ifx [1]{%
 \ifx #1\expandafter \@firstoftwo
 \else \expandafter \@secondoftwo
 \fi
}%
\providecommand \natexlab [1]{#1}%
\providecommand \enquote  [1]{``#1''}%
\providecommand \bibnamefont  [1]{#1}%
\providecommand \bibfnamefont [1]{#1}%
\providecommand \citenamefont [1]{#1}%
\providecommand \href@noop [0]{\@secondoftwo}%
\providecommand \href [0]{\begingroup \@sanitize@url \@href}%
\providecommand \@href[1]{\@@startlink{#1}\@@href}%
\providecommand \@@href[1]{\endgroup#1\@@endlink}%
\providecommand \@sanitize@url [0]{\catcode `\\12\catcode `\$12\catcode
  `\&12\catcode `\#12\catcode `\^12\catcode `\_12\catcode `\%12\relax}%
\providecommand \@@startlink[1]{}%
\providecommand \@@endlink[0]{}%
\providecommand \url  [0]{\begingroup\@sanitize@url \@url }%
\providecommand \@url [1]{\endgroup\@href {#1}{\urlprefix }}%
\providecommand \urlprefix  [0]{URL }%
\providecommand \Eprint [0]{\href }%
\providecommand \doibase [0]{http://dx.doi.org/}%
\providecommand \selectlanguage [0]{\@gobble}%
\providecommand \bibinfo  [0]{\@secondoftwo}%
\providecommand \bibfield  [0]{\@secondoftwo}%
\providecommand \translation [1]{[#1]}%
\providecommand \BibitemOpen [0]{}%
\providecommand \bibitemStop [0]{}%
\providecommand \bibitemNoStop [0]{.\EOS\space}%
\providecommand \EOS [0]{\spacefactor3000\relax}%
\providecommand \BibitemShut  [1]{\csname bibitem#1\endcsname}%
\let\auto@bib@innerbib\@empty
\bibitem [{\citenamefont {Bersuker}(2013)}]{bersuker2013jahn}%
  \BibitemOpen
  \bibfield  {author} {\bibinfo {author} {\bibfnamefont {I.~B.}\ \bibnamefont
  {Bersuker}},\ }\href@noop {} {\emph {\bibinfo {title} {The Jahn\,--\,Teller
  effect and vibronic interactions in modern chemistry}}}\ (\bibinfo
  {publisher} {Springer Science \& Business Media},\ \bibinfo {year}
  {2013})\BibitemShut {NoStop}%
\bibitem [{\citenamefont {Kimura}\ \emph {et~al.}(2008)\citenamefont {Kimura},
  \citenamefont {Sekio}, \citenamefont {Nakamura}, \citenamefont {Siegrist},\
  and\ \citenamefont {Ramirez}}]{kimura2008cupric}%
  \BibitemOpen
  \bibfield  {author} {\bibinfo {author} {\bibfnamefont {T.}~\bibnamefont
  {Kimura}}, \bibinfo {author} {\bibfnamefont {Y.}~\bibnamefont {Sekio}},
  \bibinfo {author} {\bibfnamefont {H.}~\bibnamefont {Nakamura}}, \bibinfo
  {author} {\bibfnamefont {T.}~\bibnamefont {Siegrist}}, \ and\ \bibinfo
  {author} {\bibfnamefont {A.~P.}\ \bibnamefont {Ramirez}},\ }\href {\doibase
  10.1038/nmat2125} {\bibfield  {journal} {\bibinfo  {journal} {Nat Mater}\
  }\textbf {\bibinfo {volume} {7}},\ \bibinfo {pages} {291} (\bibinfo {year}
  {2008})}\BibitemShut {NoStop}%
\bibitem [{\citenamefont {Kazimierczuk}\ \emph {et~al.}(2014)\citenamefont
  {Kazimierczuk}, \citenamefont {Fr{\"o}hlich}, \citenamefont {Scheel},
  \citenamefont {Stolz},\ and\ \citenamefont {Bayer}}]{kazimierczuk2014giant}%
  \BibitemOpen
  \bibfield  {author} {\bibinfo {author} {\bibfnamefont {T.}~\bibnamefont
  {Kazimierczuk}}, \bibinfo {author} {\bibfnamefont {D.}~\bibnamefont
  {Fr{\"o}hlich}}, \bibinfo {author} {\bibfnamefont {S.}~\bibnamefont
  {Scheel}}, \bibinfo {author} {\bibfnamefont {H.}~\bibnamefont {Stolz}}, \
  and\ \bibinfo {author} {\bibfnamefont {M.}~\bibnamefont {Bayer}},\
  }\href@noop {} {\bibfield  {journal} {\bibinfo  {journal} {Nature}\ }\textbf
  {\bibinfo {volume} {514}},\ \bibinfo {pages} {343} (\bibinfo {year}
  {2014})}\BibitemShut {NoStop}%
\bibitem [{\citenamefont {Hase}\ \emph {et~al.}(1993)\citenamefont {Hase},
  \citenamefont {Terasaki},\ and\ \citenamefont
  {Uchinokura}}]{hase1993observation}%
  \BibitemOpen
  \bibfield  {author} {\bibinfo {author} {\bibfnamefont {M.}~\bibnamefont
  {Hase}}, \bibinfo {author} {\bibfnamefont {I.}~\bibnamefont {Terasaki}}, \
  and\ \bibinfo {author} {\bibfnamefont {K.}~\bibnamefont {Uchinokura}},\
  }\href {\doibase 10.1103/PhysRevLett.70.3651} {\bibfield  {journal} {\bibinfo
   {journal} {Phys. Rev. Lett.}\ }\textbf {\bibinfo {volume} {70}},\ \bibinfo
  {pages} {3651} (\bibinfo {year} {1993})}\BibitemShut {NoStop}%
\bibitem [{\citenamefont {Azuma}\ \emph {et~al.}(1994)\citenamefont {Azuma},
  \citenamefont {Hiroi}, \citenamefont {Takano}, \citenamefont {Ishida},\ and\
  \citenamefont {Kitaoka}}]{azuma1994observation}%
  \BibitemOpen
  \bibfield  {author} {\bibinfo {author} {\bibfnamefont {M.}~\bibnamefont
  {Azuma}}, \bibinfo {author} {\bibfnamefont {Z.}~\bibnamefont {Hiroi}},
  \bibinfo {author} {\bibfnamefont {M.}~\bibnamefont {Takano}}, \bibinfo
  {author} {\bibfnamefont {K.}~\bibnamefont {Ishida}}, \ and\ \bibinfo {author}
  {\bibfnamefont {Y.}~\bibnamefont {Kitaoka}},\ }\href {\doibase
  10.1103/PhysRevLett.73.3463} {\bibfield  {journal} {\bibinfo  {journal}
  {Phys. Rev. Lett.}\ }\textbf {\bibinfo {volume} {73}},\ \bibinfo {pages}
  {3463} (\bibinfo {year} {1994})}\BibitemShut {NoStop}%
\bibitem [{\citenamefont {Kageyama}\ \emph {et~al.}(1999)\citenamefont
  {Kageyama}, \citenamefont {Yoshimura}, \citenamefont {Stern}, \citenamefont
  {Mushnikov}, \citenamefont {Onizuka}, \citenamefont {Kato}, \citenamefont
  {Kosuge}, \citenamefont {Slichter}, \citenamefont {Goto},\ and\ \citenamefont
  {Ueda}}]{kageyama1999exact}%
  \BibitemOpen
  \bibfield  {author} {\bibinfo {author} {\bibfnamefont {H.}~\bibnamefont
  {Kageyama}}, \bibinfo {author} {\bibfnamefont {K.}~\bibnamefont {Yoshimura}},
  \bibinfo {author} {\bibfnamefont {R.}~\bibnamefont {Stern}}, \bibinfo
  {author} {\bibfnamefont {N.~V.}\ \bibnamefont {Mushnikov}}, \bibinfo {author}
  {\bibfnamefont {K.}~\bibnamefont {Onizuka}}, \bibinfo {author} {\bibfnamefont
  {M.}~\bibnamefont {Kato}}, \bibinfo {author} {\bibfnamefont {K.}~\bibnamefont
  {Kosuge}}, \bibinfo {author} {\bibfnamefont {C.~P.}\ \bibnamefont
  {Slichter}}, \bibinfo {author} {\bibfnamefont {T.}~\bibnamefont {Goto}}, \
  and\ \bibinfo {author} {\bibfnamefont {Y.}~\bibnamefont {Ueda}},\ }\href
  {\doibase 10.1103/PhysRevLett.82.3168} {\bibfield  {journal} {\bibinfo
  {journal} {Phys. Rev. Lett.}\ }\textbf {\bibinfo {volume} {82}},\ \bibinfo
  {pages} {3168} (\bibinfo {year} {1999})}\BibitemShut {NoStop}%
\bibitem [{\citenamefont {Kageyama}\ \emph {et~al.}(2000)\citenamefont
  {Kageyama}, \citenamefont {Nishi}, \citenamefont {Aso}, \citenamefont
  {Onizuka}, \citenamefont {Yosihama}, \citenamefont {Nukui}, \citenamefont
  {Kodama}, \citenamefont {Kakurai},\ and\ \citenamefont
  {Ueda}}]{kageyama2000direct}%
  \BibitemOpen
  \bibfield  {author} {\bibinfo {author} {\bibfnamefont {H.}~\bibnamefont
  {Kageyama}}, \bibinfo {author} {\bibfnamefont {M.}~\bibnamefont {Nishi}},
  \bibinfo {author} {\bibfnamefont {N.}~\bibnamefont {Aso}}, \bibinfo {author}
  {\bibfnamefont {K.}~\bibnamefont {Onizuka}}, \bibinfo {author} {\bibfnamefont
  {T.}~\bibnamefont {Yosihama}}, \bibinfo {author} {\bibfnamefont
  {K.}~\bibnamefont {Nukui}}, \bibinfo {author} {\bibfnamefont
  {K.}~\bibnamefont {Kodama}}, \bibinfo {author} {\bibfnamefont
  {K.}~\bibnamefont {Kakurai}}, \ and\ \bibinfo {author} {\bibfnamefont
  {Y.}~\bibnamefont {Ueda}},\ }\href {\doibase 10.1103/PhysRevLett.84.5876}
  {\bibfield  {journal} {\bibinfo  {journal} {Phys. Rev. Lett.}\ }\textbf
  {\bibinfo {volume} {84}},\ \bibinfo {pages} {5876} (\bibinfo {year}
  {2000})}\BibitemShut {NoStop}%
\bibitem [{\citenamefont {Kodama}\ \emph {et~al.}(2002)\citenamefont {Kodama},
  \citenamefont {Takigawa}, \citenamefont {Horvati{\'c}}, \citenamefont
  {Berthier}, \citenamefont {Kageyama}, \citenamefont {Ueda}, \citenamefont
  {Miyahara}, \citenamefont {Becca},\ and\ \citenamefont
  {Mila}}]{kodama2002magnetic}%
  \BibitemOpen
  \bibfield  {author} {\bibinfo {author} {\bibfnamefont {K.}~\bibnamefont
  {Kodama}}, \bibinfo {author} {\bibfnamefont {M.}~\bibnamefont {Takigawa}},
  \bibinfo {author} {\bibfnamefont {M.}~\bibnamefont {Horvati{\'c}}}, \bibinfo
  {author} {\bibfnamefont {C.}~\bibnamefont {Berthier}}, \bibinfo {author}
  {\bibfnamefont {H.}~\bibnamefont {Kageyama}}, \bibinfo {author}
  {\bibfnamefont {Y.}~\bibnamefont {Ueda}}, \bibinfo {author} {\bibfnamefont
  {S.}~\bibnamefont {Miyahara}}, \bibinfo {author} {\bibfnamefont
  {F.}~\bibnamefont {Becca}}, \ and\ \bibinfo {author} {\bibfnamefont
  {F.}~\bibnamefont {Mila}},\ }\href {\doibase 10.1126/science.1075045}
  {\bibfield  {journal} {\bibinfo  {journal} {Science}\ }\textbf {\bibinfo
  {volume} {298}},\ \bibinfo {pages} {395} (\bibinfo {year}
  {2002})}\BibitemShut {NoStop}%
\bibitem [{\citenamefont {Bert}\ \emph {et~al.}(2007)\citenamefont {Bert},
  \citenamefont {Nakamae}, \citenamefont {Ladieu}, \citenamefont {L'H\^ote},
  \citenamefont {Bonville}, \citenamefont {Duc}, \citenamefont {Trombe},\ and\
  \citenamefont {Mendels}}]{bert2007low}%
  \BibitemOpen
  \bibfield  {author} {\bibinfo {author} {\bibfnamefont {F.}~\bibnamefont
  {Bert}}, \bibinfo {author} {\bibfnamefont {S.}~\bibnamefont {Nakamae}},
  \bibinfo {author} {\bibfnamefont {F.}~\bibnamefont {Ladieu}}, \bibinfo
  {author} {\bibfnamefont {D.}~\bibnamefont {L'H\^ote}}, \bibinfo {author}
  {\bibfnamefont {P.}~\bibnamefont {Bonville}}, \bibinfo {author}
  {\bibfnamefont {F.}~\bibnamefont {Duc}}, \bibinfo {author} {\bibfnamefont
  {J.-C.}\ \bibnamefont {Trombe}}, \ and\ \bibinfo {author} {\bibfnamefont
  {P.}~\bibnamefont {Mendels}},\ }\href {\doibase 10.1103/PhysRevB.76.132411}
  {\bibfield  {journal} {\bibinfo  {journal} {Phys. Rev. B}\ }\textbf {\bibinfo
  {volume} {76}},\ \bibinfo {pages} {132411} (\bibinfo {year}
  {2007})}\BibitemShut {NoStop}%
\bibitem [{\citenamefont {Park}\ \emph {et~al.}(2007)\citenamefont {Park},
  \citenamefont {Choi}, \citenamefont {Zhang},\ and\ \citenamefont
  {Cheong}}]{park2007ferroelectricity}%
  \BibitemOpen
  \bibfield  {author} {\bibinfo {author} {\bibfnamefont {S.}~\bibnamefont
  {Park}}, \bibinfo {author} {\bibfnamefont {Y.~J.}\ \bibnamefont {Choi}},
  \bibinfo {author} {\bibfnamefont {C.~L.}\ \bibnamefont {Zhang}}, \ and\
  \bibinfo {author} {\bibfnamefont {S.-W.}\ \bibnamefont {Cheong}},\ }\href
  {\doibase 10.1103/PhysRevLett.98.057601} {\bibfield  {journal} {\bibinfo
  {journal} {Phys. Rev. Lett.}\ }\textbf {\bibinfo {volume} {98}},\ \bibinfo
  {pages} {057601} (\bibinfo {year} {2007})}\BibitemShut {NoStop}%
\bibitem [{\citenamefont {Pisarev}\ \emph {et~al.}(2006)\citenamefont
  {Pisarev}, \citenamefont {Moskvin}, \citenamefont {Kalashnikova},
  \citenamefont {Bush},\ and\ \citenamefont {Rasing}}]{pisarev2006anomalous}%
  \BibitemOpen
  \bibfield  {author} {\bibinfo {author} {\bibfnamefont {R.~V.}\ \bibnamefont
  {Pisarev}}, \bibinfo {author} {\bibfnamefont {A.~S.}\ \bibnamefont
  {Moskvin}}, \bibinfo {author} {\bibfnamefont {A.~M.}\ \bibnamefont
  {Kalashnikova}}, \bibinfo {author} {\bibfnamefont {A.~A.}\ \bibnamefont
  {Bush}}, \ and\ \bibinfo {author} {\bibfnamefont {T.}~\bibnamefont
  {Rasing}},\ }\href {\doibase 10.1103/PhysRevB.74.132509} {\bibfield
  {journal} {\bibinfo  {journal} {Phys. Rev. B}\ }\textbf {\bibinfo {volume}
  {74}},\ \bibinfo {pages} {132509} (\bibinfo {year} {2006})}\BibitemShut
  {NoStop}%
\bibitem [{\citenamefont {Podlesnyak}\ \emph {et~al.}(2016)\citenamefont
  {Podlesnyak}, \citenamefont {Anovitz}, \citenamefont {Kolesnikov},
  \citenamefont {Matsuda}, \citenamefont {Prisk}, \citenamefont {Toth},\ and\
  \citenamefont {Ehlers}}]{podlesnyak2016coupled}%
  \BibitemOpen
  \bibfield  {author} {\bibinfo {author} {\bibfnamefont {A.}~\bibnamefont
  {Podlesnyak}}, \bibinfo {author} {\bibfnamefont {L.~M.}\ \bibnamefont
  {Anovitz}}, \bibinfo {author} {\bibfnamefont {A.~I.}\ \bibnamefont
  {Kolesnikov}}, \bibinfo {author} {\bibfnamefont {M.}~\bibnamefont {Matsuda}},
  \bibinfo {author} {\bibfnamefont {T.~R.}\ \bibnamefont {Prisk}}, \bibinfo
  {author} {\bibfnamefont {S.}~\bibnamefont {Toth}}, \ and\ \bibinfo {author}
  {\bibfnamefont {G.}~\bibnamefont {Ehlers}},\ }\href {\doibase
  10.1103/PhysRevB.93.064426} {\bibfield  {journal} {\bibinfo  {journal} {Phys.
  Rev. B}\ }\textbf {\bibinfo {volume} {93}},\ \bibinfo {pages} {064426}
  (\bibinfo {year} {2016})}\BibitemShut {NoStop}%
\bibitem [{\citenamefont {Janson}\ \emph {et~al.}(2014)\citenamefont {Janson},
  \citenamefont {Rousochatzakis}, \citenamefont {Tsirlin}, \citenamefont
  {Belesi}, \citenamefont {Leonov}, \citenamefont {R{\"o}{\ss}ler},
  \citenamefont {van~den Brink},\ and\ \citenamefont
  {Rosner}}]{janson2014skyrmions}%
  \BibitemOpen
  \bibfield  {author} {\bibinfo {author} {\bibfnamefont {O.}~\bibnamefont
  {Janson}}, \bibinfo {author} {\bibfnamefont {I.}~\bibnamefont
  {Rousochatzakis}}, \bibinfo {author} {\bibfnamefont {A.~A.}\ \bibnamefont
  {Tsirlin}}, \bibinfo {author} {\bibfnamefont {M.}~\bibnamefont {Belesi}},
  \bibinfo {author} {\bibfnamefont {A.~A.}\ \bibnamefont {Leonov}}, \bibinfo
  {author} {\bibfnamefont {U.~K.}\ \bibnamefont {R{\"o}{\ss}ler}}, \bibinfo
  {author} {\bibfnamefont {J.}~\bibnamefont {van~den Brink}}, \ and\ \bibinfo
  {author} {\bibfnamefont {H.}~\bibnamefont {Rosner}},\ }\href {\doibase
  10.1038/ncomms6376} {\bibfield  {journal} {\bibinfo  {journal} {Nature
  Communications}\ }\textbf {\bibinfo {volume} {5}},\ \bibinfo {pages} {5376}
  (\bibinfo {year} {2014})}\BibitemShut {NoStop}%
\bibitem [{\citenamefont {Landee}\ and\ \citenamefont
  {Turnbull}(2013)}]{landee2013recent}%
  \BibitemOpen
  \bibfield  {author} {\bibinfo {author} {\bibfnamefont {C.~P.}\ \bibnamefont
  {Landee}}\ and\ \bibinfo {author} {\bibfnamefont {M.~M.}\ \bibnamefont
  {Turnbull}},\ }\href {\doibase 10.1002/ejic.201300133} {\bibfield  {journal}
  {\bibinfo  {journal} {European Journal of Inorganic Chemistry}\ }\textbf
  {\bibinfo {volume} {2013}},\ \bibinfo {pages} {2266} (\bibinfo {year}
  {2013})}\BibitemShut {NoStop}%
\bibitem [{\citenamefont {Keimer}\ \emph {et~al.}(2015)\citenamefont {Keimer},
  \citenamefont {Kivelson}, \citenamefont {Norman}, \citenamefont {Uchida},\
  and\ \citenamefont {Zaanen}}]{keimer2015quantum}%
  \BibitemOpen
  \bibfield  {author} {\bibinfo {author} {\bibfnamefont {B.}~\bibnamefont
  {Keimer}}, \bibinfo {author} {\bibfnamefont {S.~A.}\ \bibnamefont
  {Kivelson}}, \bibinfo {author} {\bibfnamefont {M.~R.}\ \bibnamefont
  {Norman}}, \bibinfo {author} {\bibfnamefont {S.}~\bibnamefont {Uchida}}, \
  and\ \bibinfo {author} {\bibfnamefont {J.}~\bibnamefont {Zaanen}},\ }\href
  {\doibase 10.1038/nature14165} {\bibfield  {journal} {\bibinfo  {journal}
  {Nature}\ }\textbf {\bibinfo {volume} {518}},\ \bibinfo {pages} {179}
  (\bibinfo {year} {2015})}\BibitemShut {NoStop}%
\bibitem [{\citenamefont {Martinez-Ripoll}\ \emph {et~al.}(1971)\citenamefont
  {Martinez-Ripoll}, \citenamefont {Mart{\'\i}nez-Carrera},\ and\ \citenamefont
  {Garc{\'\i}a-Blanco}}]{martinez1971crystal}%
  \BibitemOpen
  \bibfield  {author} {\bibinfo {author} {\bibfnamefont {M.}~\bibnamefont
  {Martinez-Ripoll}}, \bibinfo {author} {\bibfnamefont {S.}~\bibnamefont
  {Mart{\'\i}nez-Carrera}}, \ and\ \bibinfo {author} {\bibfnamefont
  {S.}~\bibnamefont {Garc{\'\i}a-Blanco}},\ }\href {\doibase
  10.1107/S0567740871002760} {\bibfield  {journal} {\bibinfo  {journal} {Acta
  Crystallographica Section B}\ }\textbf {\bibinfo {volume} {27}},\ \bibinfo
  {pages} {677} (\bibinfo {year} {1971})}\BibitemShut {NoStop}%
\bibitem [{\citenamefont {Boehm}\ \emph {et~al.}(2003)\citenamefont {Boehm},
  \citenamefont {Roessli}, \citenamefont {Schefer}, \citenamefont {Wills},
  \citenamefont {Ouladdiaf}, \citenamefont {Leli\`evre-Berna}, \citenamefont
  {Staub},\ and\ \citenamefont {Petrakovskii}}]{boehm2003complex}%
  \BibitemOpen
  \bibfield  {author} {\bibinfo {author} {\bibfnamefont {M.}~\bibnamefont
  {Boehm}}, \bibinfo {author} {\bibfnamefont {B.}~\bibnamefont {Roessli}},
  \bibinfo {author} {\bibfnamefont {J.}~\bibnamefont {Schefer}}, \bibinfo
  {author} {\bibfnamefont {A.~S.}\ \bibnamefont {Wills}}, \bibinfo {author}
  {\bibfnamefont {B.}~\bibnamefont {Ouladdiaf}}, \bibinfo {author}
  {\bibfnamefont {E.}~\bibnamefont {Leli\`evre-Berna}}, \bibinfo {author}
  {\bibfnamefont {U.}~\bibnamefont {Staub}}, \ and\ \bibinfo {author}
  {\bibfnamefont {G.~A.}\ \bibnamefont {Petrakovskii}},\ }\href {\doibase
  10.1103/PhysRevB.68.024405} {\bibfield  {journal} {\bibinfo  {journal} {Phys.
  Rev. B}\ }\textbf {\bibinfo {volume} {68}},\ \bibinfo {pages} {024405}
  (\bibinfo {year} {2003})}\BibitemShut {NoStop}%
\bibitem [{\citenamefont {Pisarev}\ \emph {et~al.}(2004)\citenamefont
  {Pisarev}, \citenamefont {S\"anger}, \citenamefont {Petrakovskii},\ and\
  \citenamefont {Fiebig}}]{pisarev2004magnetic}%
  \BibitemOpen
  \bibfield  {author} {\bibinfo {author} {\bibfnamefont {R.~V.}\ \bibnamefont
  {Pisarev}}, \bibinfo {author} {\bibfnamefont {I.}~\bibnamefont {S\"anger}},
  \bibinfo {author} {\bibfnamefont {G.~A.}\ \bibnamefont {Petrakovskii}}, \
  and\ \bibinfo {author} {\bibfnamefont {M.}~\bibnamefont {Fiebig}},\ }\href
  {\doibase 10.1103/PhysRevLett.93.037204} {\bibfield  {journal} {\bibinfo
  {journal} {Phys. Rev. Lett.}\ }\textbf {\bibinfo {volume} {93}},\ \bibinfo
  {pages} {037204} (\bibinfo {year} {2004})}\BibitemShut {NoStop}%
\bibitem [{\citenamefont {Pisarev}\ \emph {et~al.}(2011)\citenamefont
  {Pisarev}, \citenamefont {Kalashnikova}, \citenamefont {Sch\"ops},\ and\
  \citenamefont {Bezmaternykh}}]{pisarev2011electronic}%
  \BibitemOpen
  \bibfield  {author} {\bibinfo {author} {\bibfnamefont {R.~V.}\ \bibnamefont
  {Pisarev}}, \bibinfo {author} {\bibfnamefont {A.~M.}\ \bibnamefont
  {Kalashnikova}}, \bibinfo {author} {\bibfnamefont {O.}~\bibnamefont
  {Sch\"ops}}, \ and\ \bibinfo {author} {\bibfnamefont {L.~N.}\ \bibnamefont
  {Bezmaternykh}},\ }\href {\doibase 10.1103/PhysRevB.84.075160} {\bibfield
  {journal} {\bibinfo  {journal} {Phys. Rev. B}\ }\textbf {\bibinfo {volume}
  {84}},\ \bibinfo {pages} {075160} (\bibinfo {year} {2011})}\BibitemShut
  {NoStop}%
\bibitem [{\citenamefont {Pisarev}\ \emph {et~al.}(2013)\citenamefont
  {Pisarev}, \citenamefont {Boldyrev}, \citenamefont {Popova}, \citenamefont
  {Smirnov}, \citenamefont {Davydov}, \citenamefont {Bezmaternykh},
  \citenamefont {Smirnov},\ and\ \citenamefont
  {Kazimirov}}]{pisarev2013lattice}%
  \BibitemOpen
  \bibfield  {author} {\bibinfo {author} {\bibfnamefont {R.~V.}\ \bibnamefont
  {Pisarev}}, \bibinfo {author} {\bibfnamefont {K.~N.}\ \bibnamefont
  {Boldyrev}}, \bibinfo {author} {\bibfnamefont {M.~N.}\ \bibnamefont
  {Popova}}, \bibinfo {author} {\bibfnamefont {A.~N.}\ \bibnamefont {Smirnov}},
  \bibinfo {author} {\bibfnamefont {V.~Y.}\ \bibnamefont {Davydov}}, \bibinfo
  {author} {\bibfnamefont {L.~N.}\ \bibnamefont {Bezmaternykh}}, \bibinfo
  {author} {\bibfnamefont {M.~B.}\ \bibnamefont {Smirnov}}, \ and\ \bibinfo
  {author} {\bibfnamefont {V.~Y.}\ \bibnamefont {Kazimirov}},\ }\href {\doibase
  10.1103/PhysRevB.88.024301} {\bibfield  {journal} {\bibinfo  {journal} {Phys.
  Rev. B}\ }\textbf {\bibinfo {volume} {88}},\ \bibinfo {pages} {024301}
  (\bibinfo {year} {2013})}\BibitemShut {NoStop}%
\bibitem [{\citenamefont {Boldyrev}\ \emph {et~al.}(2015)\citenamefont
  {Boldyrev}, \citenamefont {Pisarev}, \citenamefont {Bezmaternykh},\ and\
  \citenamefont {Popova}}]{boldyrev2015antiferromagnetic}%
  \BibitemOpen
  \bibfield  {author} {\bibinfo {author} {\bibfnamefont {K.~N.}\ \bibnamefont
  {Boldyrev}}, \bibinfo {author} {\bibfnamefont {R.~V.}\ \bibnamefont
  {Pisarev}}, \bibinfo {author} {\bibfnamefont {L.~N.}\ \bibnamefont
  {Bezmaternykh}}, \ and\ \bibinfo {author} {\bibfnamefont {M.~N.}\
  \bibnamefont {Popova}},\ }\href {\doibase 10.1103/PhysRevLett.114.247210}
  {\bibfield  {journal} {\bibinfo  {journal} {Phys. Rev. Lett.}\ }\textbf
  {\bibinfo {volume} {114}},\ \bibinfo {pages} {247210} (\bibinfo {year}
  {2015})}\BibitemShut {NoStop}%
\bibitem [{\citenamefont {Depmeier}\ and\ \citenamefont
  {Schmid}(1982)}]{depmeier1982palladium}%
  \BibitemOpen
  \bibfield  {author} {\bibinfo {author} {\bibfnamefont {W.}~\bibnamefont
  {Depmeier}}\ and\ \bibinfo {author} {\bibfnamefont {H.}~\bibnamefont
  {Schmid}},\ }\href {\doibase 10.1107/S0567740882003495} {\bibfield  {journal}
  {\bibinfo  {journal} {Acta Crystallographica Section B}\ }\textbf {\bibinfo
  {volume} {38}},\ \bibinfo {pages} {605} (\bibinfo {year} {1982})}\BibitemShut
  {NoStop}%
\bibitem [{\citenamefont {Behm}(1982)}]{behm1982pentadecacopper}%
  \BibitemOpen
  \bibfield  {author} {\bibinfo {author} {\bibfnamefont {H.}~\bibnamefont
  {Behm}},\ }\href {\doibase 10.1107/S0567740882009947} {\bibfield  {journal}
  {\bibinfo  {journal} {Acta Crystallographica Section B}\ }\textbf {\bibinfo
  {volume} {38}},\ \bibinfo {pages} {2781} (\bibinfo {year}
  {1982})}\BibitemShut {NoStop}%
\bibitem [{\citenamefont {Kuratieva}\ \emph {et~al.}(2009)\citenamefont
  {Kuratieva}, \citenamefont {Mikhailova},\ and\ \citenamefont
  {Ehrenberg}}]{kuratieva2009new}%
  \BibitemOpen
  \bibfield  {author} {\bibinfo {author} {\bibfnamefont {N.~V.}\ \bibnamefont
  {Kuratieva}}, \bibinfo {author} {\bibfnamefont {D.}~\bibnamefont
  {Mikhailova}}, \ and\ \bibinfo {author} {\bibfnamefont {H.}~\bibnamefont
  {Ehrenberg}},\ }\href {\doibase 10.1107/S0108270109036865} {\bibfield
  {journal} {\bibinfo  {journal} {Acta Crystallographica Section C}\ }\textbf
  {\bibinfo {volume} {65}},\ \bibinfo {pages} {i85} (\bibinfo {year}
  {2009})}\BibitemShut {NoStop}%
\bibitem [{\citenamefont {Zhang}\ \emph {et~al.}(2017)\citenamefont {Zhang},
  \citenamefont {Zhao}, \citenamefont {Zhang}, \citenamefont {Ma},
  \citenamefont {Xin},\ and\ \citenamefont {Li}}]{zhang2017synthesis}%
  \BibitemOpen
  \bibfield  {author} {\bibinfo {author} {\bibfnamefont {R.-H.}\ \bibnamefont
  {Zhang}}, \bibinfo {author} {\bibfnamefont {D.}~\bibnamefont {Zhao}},
  \bibinfo {author} {\bibfnamefont {L.}~\bibnamefont {Zhang}}, \bibinfo
  {author} {\bibfnamefont {F.-X.}\ \bibnamefont {Ma}}, \bibinfo {author}
  {\bibfnamefont {X.}~\bibnamefont {Xin}}, \ and\ \bibinfo {author}
  {\bibfnamefont {F.-F.}\ \bibnamefont {Li}},\ }\href {\doibase
  10.1080/15533174.2016.1186085} {\bibfield  {journal} {\bibinfo  {journal}
  {Inorganic and Nano-Metal Chemistry}\ }\textbf {\bibinfo {volume} {47}},\
  \bibinfo {pages} {521} (\bibinfo {year} {2017})}\BibitemShut {NoStop}%
\bibitem [{\citenamefont {Petrakovskii}\ \emph {et~al.}(1999)\citenamefont
  {Petrakovskii}, \citenamefont {Sablina}, \citenamefont {Vorotynov},
  \citenamefont {Bayukov}, \citenamefont {Bovina}, \citenamefont {Bondarenko},
  \citenamefont {Szymczak}, \citenamefont {Baran},\ and\ \citenamefont
  {Szymczak}}]{petrakovskii1999synthesis}%
  \BibitemOpen
  \bibfield  {author} {\bibinfo {author} {\bibfnamefont {G.~A.}\ \bibnamefont
  {Petrakovskii}}, \bibinfo {author} {\bibfnamefont {K.~A.}\ \bibnamefont
  {Sablina}}, \bibinfo {author} {\bibfnamefont {A.~M.}\ \bibnamefont
  {Vorotynov}}, \bibinfo {author} {\bibfnamefont {O.~A.}\ \bibnamefont
  {Bayukov}}, \bibinfo {author} {\bibfnamefont {A.~F.}\ \bibnamefont {Bovina}},
  \bibinfo {author} {\bibfnamefont {G.~V.}\ \bibnamefont {Bondarenko}},
  \bibinfo {author} {\bibfnamefont {R.}~\bibnamefont {Szymczak}}, \bibinfo
  {author} {\bibfnamefont {M.}~\bibnamefont {Baran}}, \ and\ \bibinfo {author}
  {\bibfnamefont {H.}~\bibnamefont {Szymczak}},\ }\href {\doibase
  10.1134/1.1130835} {\bibfield  {journal} {\bibinfo  {journal} {Physics of the
  Solid State}\ }\textbf {\bibinfo {volume} {41}},\ \bibinfo {pages} {610}
  (\bibinfo {year} {1999})}\BibitemShut {NoStop}%
\bibitem [{\citenamefont {Petrakovski{\u{\i}}}\ \emph
  {et~al.}(2007)\citenamefont {Petrakovski{\u{\i}}}, \citenamefont
  {Bezmaternykh}, \citenamefont {Bayukov}, \citenamefont {Popov}, \citenamefont
  {Schefer}, \citenamefont {Neidermayer}, \citenamefont {Aleshkevich},\ and\
  \citenamefont {Szymczak}}]{petrakovskiui2007magnetic}%
  \BibitemOpen
  \bibfield  {author} {\bibinfo {author} {\bibfnamefont {G.~A.}\ \bibnamefont
  {Petrakovski{\u{\i}}}}, \bibinfo {author} {\bibfnamefont {L.~N.}\
  \bibnamefont {Bezmaternykh}}, \bibinfo {author} {\bibfnamefont {O.~A.}\
  \bibnamefont {Bayukov}}, \bibinfo {author} {\bibfnamefont {M.~A.}\
  \bibnamefont {Popov}}, \bibinfo {author} {\bibfnamefont {J.}~\bibnamefont
  {Schefer}}, \bibinfo {author} {\bibfnamefont {C.}~\bibnamefont
  {Neidermayer}}, \bibinfo {author} {\bibfnamefont {P.}~\bibnamefont
  {Aleshkevich}}, \ and\ \bibinfo {author} {\bibfnamefont {R.}~\bibnamefont
  {Szymczak}},\ }\href {\doibase 10.1134/S1063783407070207} {\bibfield
  {journal} {\bibinfo  {journal} {Physics of the Solid State}\ }\textbf
  {\bibinfo {volume} {49}},\ \bibinfo {pages} {1315} (\bibinfo {year}
  {2007})}\BibitemShut {NoStop}%
\bibitem [{\citenamefont {Kudo}\ \emph {et~al.}(2003)\citenamefont {Kudo},
  \citenamefont {Noji}, \citenamefont {Koike}, \citenamefont {Sakon},
  \citenamefont {Motokawa}, \citenamefont {Nishizaki},\ and\ \citenamefont
  {Kobayashi}}]{kudo2003anisotropic}%
  \BibitemOpen
  \bibfield  {author} {\bibinfo {author} {\bibfnamefont {K.}~\bibnamefont
  {Kudo}}, \bibinfo {author} {\bibfnamefont {T.}~\bibnamefont {Noji}}, \bibinfo
  {author} {\bibfnamefont {Y.}~\bibnamefont {Koike}}, \bibinfo {author}
  {\bibfnamefont {T.}~\bibnamefont {Sakon}}, \bibinfo {author} {\bibfnamefont
  {M.}~\bibnamefont {Motokawa}}, \bibinfo {author} {\bibfnamefont
  {T.}~\bibnamefont {Nishizaki}}, \ and\ \bibinfo {author} {\bibfnamefont
  {N.}~\bibnamefont {Kobayashi}},\ }\href {\doibase 10.1143/jpsj.72.569}
  {\bibfield  {journal} {\bibinfo  {journal} {Journal of the Physical Society
  of Japan}\ }\textbf {\bibinfo {volume} {72}},\ \bibinfo {pages} {569}
  (\bibinfo {year} {2003})}\BibitemShut {NoStop}%
\bibitem [{\citenamefont {Kudo}\ \emph {et~al.}(2001)\citenamefont {Kudo},
  \citenamefont {Noji},\ and\ \citenamefont
  {Koike}}]{kudo2001antiferromagnetic}%
  \BibitemOpen
  \bibfield  {author} {\bibinfo {author} {\bibfnamefont {K.}~\bibnamefont
  {Kudo}}, \bibinfo {author} {\bibfnamefont {T.}~\bibnamefont {Noji}}, \ and\
  \bibinfo {author} {\bibfnamefont {Y.}~\bibnamefont {Koike}},\ }\href
  {\doibase 10.1143/jpsj.70.935} {\bibfield  {journal} {\bibinfo  {journal}
  {Journal of the Physical Society of Japan}\ }\textbf {\bibinfo {volume}
  {70}},\ \bibinfo {pages} {935} (\bibinfo {year} {2001})}\BibitemShut
  {NoStop}%
\bibitem [{\citenamefont {Balaev}\ \emph {et~al.}(2016)\citenamefont {Balaev},
  \citenamefont {Sablina}, \citenamefont {Freydman}, \citenamefont {Krasikov},\
  and\ \citenamefont {Bovina}}]{balaev2016interrelation}%
  \BibitemOpen
  \bibfield  {author} {\bibinfo {author} {\bibfnamefont {D.~A.}\ \bibnamefont
  {Balaev}}, \bibinfo {author} {\bibfnamefont {K.~A.}\ \bibnamefont {Sablina}},
  \bibinfo {author} {\bibfnamefont {A.~L.}\ \bibnamefont {Freydman}}, \bibinfo
  {author} {\bibfnamefont {A.~A.}\ \bibnamefont {Krasikov}}, \ and\ \bibinfo
  {author} {\bibfnamefont {A.~F.}\ \bibnamefont {Bovina}},\ }\href {\doibase
  10.1134/S1063783416020062} {\bibfield  {journal} {\bibinfo  {journal}
  {Physics of the Solid State}\ }\textbf {\bibinfo {volume} {58}},\ \bibinfo
  {pages} {284} (\bibinfo {year} {2016})}\BibitemShut {NoStop}%
\bibitem [{\citenamefont {Sakurai}\ \emph {et~al.}(2002)\citenamefont
  {Sakurai}, \citenamefont {Tsuboi}, \citenamefont {Kato}, \citenamefont
  {Yoshimura}, \citenamefont {Kosuge}, \citenamefont {Mitsuda}, \citenamefont
  {Mitamura},\ and\ \citenamefont {Goto}}]{sakurai2002antiferromagnetic}%
  \BibitemOpen
  \bibfield  {author} {\bibinfo {author} {\bibfnamefont {H.}~\bibnamefont
  {Sakurai}}, \bibinfo {author} {\bibfnamefont {N.}~\bibnamefont {Tsuboi}},
  \bibinfo {author} {\bibfnamefont {M.}~\bibnamefont {Kato}}, \bibinfo {author}
  {\bibfnamefont {K.}~\bibnamefont {Yoshimura}}, \bibinfo {author}
  {\bibfnamefont {K.}~\bibnamefont {Kosuge}}, \bibinfo {author} {\bibfnamefont
  {A.}~\bibnamefont {Mitsuda}}, \bibinfo {author} {\bibfnamefont
  {H.}~\bibnamefont {Mitamura}}, \ and\ \bibinfo {author} {\bibfnamefont
  {T.}~\bibnamefont {Goto}},\ }\href {\doibase 10.1103/PhysRevB.66.024428}
  {\bibfield  {journal} {\bibinfo  {journal} {Phys. Rev. B}\ }\textbf {\bibinfo
  {volume} {66}},\ \bibinfo {pages} {024428} (\bibinfo {year}
  {2002})}\BibitemShut {NoStop}%
\bibitem [{\citenamefont {Fukaya}\ \emph {et~al.}(2001)\citenamefont {Fukaya},
  \citenamefont {Watanabe},\ and\ \citenamefont {Nagamine}}]{fukaya2001long}%
  \BibitemOpen
  \bibfield  {author} {\bibinfo {author} {\bibfnamefont {A.}~\bibnamefont
  {Fukaya}}, \bibinfo {author} {\bibfnamefont {I.}~\bibnamefont {Watanabe}}, \
  and\ \bibinfo {author} {\bibfnamefont {K.}~\bibnamefont {Nagamine}},\ }\href
  {\doibase 10.1143/JPSJ.70.2868} {\bibfield  {journal} {\bibinfo  {journal}
  {Journal of the Physical Society of Japan}\ }\textbf {\bibinfo {volume}
  {70}},\ \bibinfo {pages} {2868} (\bibinfo {year} {2001})}\BibitemShut
  {NoStop}%
\bibitem [{\citenamefont {Liu}\ \emph {et~al.}(2013)\citenamefont {Liu},
  \citenamefont {Wen}, \citenamefont {Zou}, \citenamefont {Zuo}, \citenamefont
  {Beran},\ and\ \citenamefont {Feng}}]{liu2013visible}%
  \BibitemOpen
  \bibfield  {author} {\bibinfo {author} {\bibfnamefont {J.}~\bibnamefont
  {Liu}}, \bibinfo {author} {\bibfnamefont {S.}~\bibnamefont {Wen}}, \bibinfo
  {author} {\bibfnamefont {X.}~\bibnamefont {Zou}}, \bibinfo {author}
  {\bibfnamefont {F.}~\bibnamefont {Zuo}}, \bibinfo {author} {\bibfnamefont
  {G.~J.~O.}\ \bibnamefont {Beran}}, \ and\ \bibinfo {author} {\bibfnamefont
  {P.}~\bibnamefont {Feng}},\ }\href {\doibase 10.1039/C2TA00522K} {\bibfield
  {journal} {\bibinfo  {journal} {J. Mater. Chem. A}\ }\textbf {\bibinfo
  {volume} {1}},\ \bibinfo {pages} {1553} (\bibinfo {year} {2013})}\BibitemShut
  {NoStop}%
\bibitem [{\citenamefont {Momma}\ and\ \citenamefont
  {Izumi}(2011)}]{momma2011vesta}%
  \BibitemOpen
  \bibfield  {author} {\bibinfo {author} {\bibfnamefont {K.}~\bibnamefont
  {Momma}}\ and\ \bibinfo {author} {\bibfnamefont {F.}~\bibnamefont {Izumi}},\
  }\href {\doibase 10.1107/S0021889811038970} {\bibfield  {journal} {\bibinfo
  {journal} {Journal of Applied Crystallography}\ }\textbf {\bibinfo {volume}
  {44}},\ \bibinfo {pages} {1272} (\bibinfo {year} {2011})}\BibitemShut
  {NoStop}%
\bibitem [{\citenamefont {Harrick}(1960)}]{Harrick1960ATR}%
  \BibitemOpen
  \bibfield  {author} {\bibinfo {author} {\bibfnamefont {N.~J.}\ \bibnamefont
  {Harrick}},\ }\href {\doibase 10.1021/j100838a005} {\bibfield  {journal}
  {\bibinfo  {journal} {J. Phys. Chem}\ }\textbf {\bibinfo {volume} {64}},\
  \bibinfo {pages} {1110} (\bibinfo {year} {1960})}\BibitemShut {NoStop}%
\bibitem [{\citenamefont {Fahrenfort}(1961)}]{Fahrenfort1961ATR}%
  \BibitemOpen
  \bibfield  {author} {\bibinfo {author} {\bibfnamefont {J.}~\bibnamefont
  {Fahrenfort}},\ }\href {\doibase 10.1016/0371-1951(61)80136-7} {\bibfield
  {journal} {\bibinfo  {journal} {Spectrochimica Acta}\ }\textbf {\bibinfo
  {volume} {17}},\ \bibinfo {pages} {698} (\bibinfo {year} {1961})}\BibitemShut
  {NoStop}%
\bibitem [{\citenamefont {Nakamoto}(2008)}]{nakamoto}%
  \BibitemOpen
  \bibfield  {author} {\bibinfo {author} {\bibfnamefont {K.}~\bibnamefont
  {Nakamoto}},\ }\enquote {\bibinfo {title} {Applications in inorganic
  chemistry},}\ in\ \href {\doibase 10.1002/9780470405840.ch2} {\emph {\bibinfo
  {booktitle} {Infrared and Raman Spectra of Inorganic and Coordination
  Compounds}}}\ (\bibinfo  {publisher} {John Wiley \& Sons, Inc.},\ \bibinfo
  {year} {2008})\ pp.\ \bibinfo {pages} {149--354}\BibitemShut {NoStop}%
\bibitem [{\citenamefont {Fausti}\ \emph {et~al.}(2006)\citenamefont {Fausti},
  \citenamefont {Nugroho}, \citenamefont {van Loosdrecht}, \citenamefont
  {Klimin}, \citenamefont {Popova},\ and\ \citenamefont
  {Bezmaternykh}}]{fausti2006raman}%
  \BibitemOpen
  \bibfield  {author} {\bibinfo {author} {\bibfnamefont {D.}~\bibnamefont
  {Fausti}}, \bibinfo {author} {\bibfnamefont {A.~A.}\ \bibnamefont {Nugroho}},
  \bibinfo {author} {\bibfnamefont {P.~H.~M.}\ \bibnamefont {van Loosdrecht}},
  \bibinfo {author} {\bibfnamefont {S.~A.}\ \bibnamefont {Klimin}}, \bibinfo
  {author} {\bibfnamefont {M.~N.}\ \bibnamefont {Popova}}, \ and\ \bibinfo
  {author} {\bibfnamefont {L.~N.}\ \bibnamefont {Bezmaternykh}},\ }\href
  {\doibase 10.1103/PhysRevB.74.024403} {\bibfield  {journal} {\bibinfo
  {journal} {Phys. Rev. B}\ }\textbf {\bibinfo {volume} {74}},\ \bibinfo
  {pages} {024403} (\bibinfo {year} {2006})}\BibitemShut {NoStop}%
\bibitem [{\citenamefont {Kitaev}\ \emph {et~al.}(1994)\citenamefont {Kitaev},
  \citenamefont {Limonov}, \citenamefont {Panfilov}, \citenamefont
  {Mirgorodskij},\ and\ \citenamefont {Evarestov}}]{kitaev1994quasi}%
  \BibitemOpen
  \bibfield  {author} {\bibinfo {author} {\bibfnamefont {Y.~E.}\ \bibnamefont
  {Kitaev}}, \bibinfo {author} {\bibfnamefont {M.~F.}\ \bibnamefont {Limonov}},
  \bibinfo {author} {\bibfnamefont {A.~G.}\ \bibnamefont {Panfilov}}, \bibinfo
  {author} {\bibfnamefont {A.~P.}\ \bibnamefont {Mirgorodskij}}, \ and\
  \bibinfo {author} {\bibfnamefont {R.~A.}\ \bibnamefont {Evarestov}},\ }\href
  {http://journals.ioffe.ru/articles/16429} {\bibfield  {journal} {\bibinfo
  {journal} {Fizika Tverdogo Tela}\ }\textbf {\bibinfo {volume} {36}},\
  \bibinfo {pages} {865} (\bibinfo {year} {1994})}\BibitemShut {NoStop}%
\bibitem [{\citenamefont {Kuz'menko}\ \emph {et~al.}(2001)\citenamefont
  {Kuz'menko}, \citenamefont {van~der Marel}, \citenamefont {van Bentum},
  \citenamefont {Tishchenko}, \citenamefont {Presura},\ and\ \citenamefont
  {Bush}}]{kuz2001infrared}%
  \BibitemOpen
  \bibfield  {author} {\bibinfo {author} {\bibfnamefont {A.~B.}\ \bibnamefont
  {Kuz'menko}}, \bibinfo {author} {\bibfnamefont {D.}~\bibnamefont {van~der
  Marel}}, \bibinfo {author} {\bibfnamefont {P.~J.~M.}\ \bibnamefont {van
  Bentum}}, \bibinfo {author} {\bibfnamefont {E.~A.}\ \bibnamefont
  {Tishchenko}}, \bibinfo {author} {\bibfnamefont {C.}~\bibnamefont {Presura}},
  \ and\ \bibinfo {author} {\bibfnamefont {A.~A.}\ \bibnamefont {Bush}},\
  }\href {\doibase 10.1103/PhysRevB.63.094303} {\bibfield  {journal} {\bibinfo
  {journal} {Phys. Rev. B}\ }\textbf {\bibinfo {volume} {63}},\ \bibinfo
  {pages} {094303} (\bibinfo {year} {2001})}\BibitemShut {NoStop}%
\bibitem [{\citenamefont {Moskvin}(1993)}]{moskvin1993pseudo}%
  \BibitemOpen
  \bibfield  {author} {\bibinfo {author} {\bibfnamefont {A.~S.}\ \bibnamefont
  {Moskvin}},\ }\href@noop {} {\bibfield  {journal} {\bibinfo  {journal} {JETP
  Lett.}\ }\textbf {\bibinfo {volume} {58}},\ \bibinfo {pages} {345} (\bibinfo
  {year} {1993})}\BibitemShut {NoStop}%
\bibitem [{\citenamefont {Moskvin}\ \emph {et~al.}(1994)\citenamefont
  {Moskvin}, \citenamefont {Loshkareva}, \citenamefont {Sukhorukov},
  \citenamefont {Sidorov},\ and\ \citenamefont
  {Samokhvalov}}]{moskvin1994characteristics}%
  \BibitemOpen
  \bibfield  {author} {\bibinfo {author} {\bibfnamefont {A.~S.}\ \bibnamefont
  {Moskvin}}, \bibinfo {author} {\bibfnamefont {N.~N.}\ \bibnamefont
  {Loshkareva}}, \bibinfo {author} {\bibfnamefont {Y.~P.}\ \bibnamefont
  {Sukhorukov}}, \bibinfo {author} {\bibfnamefont {M.~A.}\ \bibnamefont
  {Sidorov}}, \ and\ \bibinfo {author} {\bibfnamefont {A.~A.}\ \bibnamefont
  {Samokhvalov}},\ }\href@noop {} {\bibfield  {journal} {\bibinfo  {journal}
  {JETP}\ }\textbf {\bibinfo {volume} {105}},\ \bibinfo {pages} {967} (\bibinfo
  {year} {1994})}\BibitemShut {NoStop}%
\bibitem [{\citenamefont {Popova}\ \emph {et~al.}(1997)\citenamefont {Popova},
  \citenamefont {Sushkov}, \citenamefont {Vasil'ev}, \citenamefont {Isobe},\
  and\ \citenamefont {Ueda}}]{popova1997appearance}%
  \BibitemOpen
  \bibfield  {author} {\bibinfo {author} {\bibfnamefont {M.~N.}\ \bibnamefont
  {Popova}}, \bibinfo {author} {\bibfnamefont {A.~B.}\ \bibnamefont {Sushkov}},
  \bibinfo {author} {\bibfnamefont {A.~N.}\ \bibnamefont {Vasil'ev}}, \bibinfo
  {author} {\bibfnamefont {M.}~\bibnamefont {Isobe}}, \ and\ \bibinfo {author}
  {\bibfnamefont {Y.}~\bibnamefont {Ueda}},\ }\href@noop {} {\bibfield
  {journal} {\bibinfo  {journal} {JETP Letters}\ }\textbf {\bibinfo {volume}
  {65}},\ \bibinfo {pages} {743} (\bibinfo {year} {1997})}\BibitemShut
  {NoStop}%
\bibitem [{\citenamefont {de~Jongh}\ and\ \citenamefont
  {Miedema}(1974)}]{de1974experiments}%
  \BibitemOpen
  \bibfield  {author} {\bibinfo {author} {\bibfnamefont {L.~J.}\ \bibnamefont
  {de~Jongh}}\ and\ \bibinfo {author} {\bibfnamefont {A.~R.}\ \bibnamefont
  {Miedema}},\ }\href {\doibase 10.1080/00018739700101558} {\bibfield
  {journal} {\bibinfo  {journal} {Advances in Physics}\ }\textbf {\bibinfo
  {volume} {23}},\ \bibinfo {pages} {1} (\bibinfo {year} {1974})}\BibitemShut
  {NoStop}%
\bibitem [{\citenamefont {Klimin}\ \emph {et~al.}(2016)\citenamefont {Klimin},
  \citenamefont {Kuzmenko}, \citenamefont {Kashchenko},\ and\ \citenamefont
  {Popova}}]{klimin2016infrared}%
  \BibitemOpen
  \bibfield  {author} {\bibinfo {author} {\bibfnamefont {S.~A.}\ \bibnamefont
  {Klimin}}, \bibinfo {author} {\bibfnamefont {A.~B.}\ \bibnamefont
  {Kuzmenko}}, \bibinfo {author} {\bibfnamefont {M.~A.}\ \bibnamefont
  {Kashchenko}}, \ and\ \bibinfo {author} {\bibfnamefont {M.~N.}\ \bibnamefont
  {Popova}},\ }\href {\doibase 10.1103/PhysRevB.93.054304} {\bibfield
  {journal} {\bibinfo  {journal} {Phys. Rev. B}\ }\textbf {\bibinfo {volume}
  {93}},\ \bibinfo {pages} {054304} (\bibinfo {year} {2016})}\BibitemShut
  {NoStop}%
\bibitem [{\citenamefont {Wojdyr}(2010)}]{Wojdyr2010fitik}%
  \BibitemOpen
  \bibfield  {author} {\bibinfo {author} {\bibfnamefont {M.}~\bibnamefont
  {Wojdyr}},\ }\href {\doibase 10.1107/S0021889810030499} {\bibfield  {journal}
  {\bibinfo  {journal} {Journal of Applied Crystallography}\ }\textbf {\bibinfo
  {volume} {43}},\ \bibinfo {pages} {1126} (\bibinfo {year}
  {2010})}\BibitemShut {NoStop}%
\bibitem [{\citenamefont {Bues}\ \emph {et~al.}(1966)\citenamefont {Bues},
  \citenamefont {F{\"o}rster},\ and\ \citenamefont
  {Schmitt}}]{bues1966strukturen}%
  \BibitemOpen
  \bibfield  {author} {\bibinfo {author} {\bibfnamefont {W.}~\bibnamefont
  {Bues}}, \bibinfo {author} {\bibfnamefont {G.}~\bibnamefont {F{\"o}rster}}, \
  and\ \bibinfo {author} {\bibfnamefont {R.}~\bibnamefont {Schmitt}},\ }\href
  {\doibase 10.1002/zaac.19663440305} {\bibfield  {journal} {\bibinfo
  {journal} {Zeitschrift f{\"u}r anorganische und allgemeine Chemie}\ }\textbf
  {\bibinfo {volume} {344}},\ \bibinfo {pages} {148} (\bibinfo {year}
  {1966})}\BibitemShut {NoStop}%
\bibitem [{\citenamefont {Ivanov}\ \emph {et~al.}(2013)\citenamefont {Ivanov},
  \citenamefont {Abrashev}, \citenamefont {Todorov}, \citenamefont {Tomov},
  \citenamefont {Nikolova}, \citenamefont {Litvinchuk},\ and\ \citenamefont
  {Iliev}}]{ivanov2013ramancub2o4}%
  \BibitemOpen
  \bibfield  {author} {\bibinfo {author} {\bibfnamefont {V.~G.}\ \bibnamefont
  {Ivanov}}, \bibinfo {author} {\bibfnamefont {M.~V.}\ \bibnamefont
  {Abrashev}}, \bibinfo {author} {\bibfnamefont {N.~D.}\ \bibnamefont
  {Todorov}}, \bibinfo {author} {\bibfnamefont {V.}~\bibnamefont {Tomov}},
  \bibinfo {author} {\bibfnamefont {R.~P.}\ \bibnamefont {Nikolova}}, \bibinfo
  {author} {\bibfnamefont {A.~P.}\ \bibnamefont {Litvinchuk}}, \ and\ \bibinfo
  {author} {\bibfnamefont {M.~N.}\ \bibnamefont {Iliev}},\ }\href {\doibase
  10.1103/PhysRevB.88.094301} {\bibfield  {journal} {\bibinfo  {journal} {Phys.
  Rev. B}\ }\textbf {\bibinfo {volume} {88}},\ \bibinfo {pages} {094301}
  (\bibinfo {year} {2013})}\BibitemShut {NoStop}%
\bibitem [{\citenamefont {Ernst}\ \emph {et~al.}(1998)\citenamefont {Ernst},
  \citenamefont {Broholm}, \citenamefont {Kowach},\ and\ \citenamefont
  {Ramirez}}]{ernst1998phonon}%
  \BibitemOpen
  \bibfield  {author} {\bibinfo {author} {\bibfnamefont {G.}~\bibnamefont
  {Ernst}}, \bibinfo {author} {\bibfnamefont {C.}~\bibnamefont {Broholm}},
  \bibinfo {author} {\bibfnamefont {G.~R.}\ \bibnamefont {Kowach}}, \ and\
  \bibinfo {author} {\bibfnamefont {A.~P.}\ \bibnamefont {Ramirez}},\ }\href
  {\doibase 10.1038/24115} {\bibfield  {journal} {\bibinfo  {journal} {Nature}\
  }\textbf {\bibinfo {volume} {396}},\ \bibinfo {pages} {147} (\bibinfo {year}
  {1998})}\BibitemShut {NoStop}%
\bibitem [{\citenamefont {Hlinka}\ \emph {et~al.}(2014)\citenamefont {Hlinka},
  \citenamefont {Ostapchuk}, \citenamefont {Buixaderas}, \citenamefont
  {Kadlec}, \citenamefont {Kuzel}, \citenamefont {Gregora}, \citenamefont
  {Kroupa}, \citenamefont {Savinov}, \citenamefont {Klic}, \citenamefont
  {Drahokoupil}, \citenamefont {Etxebarria},\ and\ \citenamefont
  {Dec}}]{hlinka2014multiple}%
  \BibitemOpen
  \bibfield  {author} {\bibinfo {author} {\bibfnamefont {J.}~\bibnamefont
  {Hlinka}}, \bibinfo {author} {\bibfnamefont {T.}~\bibnamefont {Ostapchuk}},
  \bibinfo {author} {\bibfnamefont {E.}~\bibnamefont {Buixaderas}}, \bibinfo
  {author} {\bibfnamefont {C.}~\bibnamefont {Kadlec}}, \bibinfo {author}
  {\bibfnamefont {P.}~\bibnamefont {Kuzel}}, \bibinfo {author} {\bibfnamefont
  {I.}~\bibnamefont {Gregora}}, \bibinfo {author} {\bibfnamefont
  {J.}~\bibnamefont {Kroupa}}, \bibinfo {author} {\bibfnamefont
  {M.}~\bibnamefont {Savinov}}, \bibinfo {author} {\bibfnamefont
  {A.}~\bibnamefont {Klic}}, \bibinfo {author} {\bibfnamefont {J.}~\bibnamefont
  {Drahokoupil}}, \bibinfo {author} {\bibfnamefont {I.}~\bibnamefont
  {Etxebarria}}, \ and\ \bibinfo {author} {\bibfnamefont {J.}~\bibnamefont
  {Dec}},\ }\href {\doibase 10.1103/PhysRevLett.112.197601} {\bibfield
  {journal} {\bibinfo  {journal} {Phys. Rev. Lett.}\ }\textbf {\bibinfo
  {volume} {112}},\ \bibinfo {pages} {197601} (\bibinfo {year}
  {2014})}\BibitemShut {NoStop}%
\bibitem [{\citenamefont {Brandt}\ \emph {et~al.}(2008)\citenamefont {Brandt},
  \citenamefont {Chikishev}, \citenamefont {Kargovsky}, \citenamefont
  {Nazarov}, \citenamefont {Parashchuk}, \citenamefont {Sapozhnikov},
  \citenamefont {Smirnova}, \citenamefont {Shkurinov},\ and\ \citenamefont
  {Sumbatyan}}]{brandt2008terahertz}%
  \BibitemOpen
  \bibfield  {author} {\bibinfo {author} {\bibfnamefont {N.~N.}\ \bibnamefont
  {Brandt}}, \bibinfo {author} {\bibfnamefont {A.~Y.}\ \bibnamefont
  {Chikishev}}, \bibinfo {author} {\bibfnamefont {A.~V.}\ \bibnamefont
  {Kargovsky}}, \bibinfo {author} {\bibfnamefont {M.~M.}\ \bibnamefont
  {Nazarov}}, \bibinfo {author} {\bibfnamefont {O.}~\bibnamefont {Parashchuk}},
  \bibinfo {author} {\bibfnamefont {D.~A.}\ \bibnamefont {Sapozhnikov}},
  \bibinfo {author} {\bibfnamefont {I.~N.}\ \bibnamefont {Smirnova}}, \bibinfo
  {author} {\bibfnamefont {A.~P.}\ \bibnamefont {Shkurinov}}, \ and\ \bibinfo
  {author} {\bibfnamefont {N.~V.}\ \bibnamefont {Sumbatyan}},\ }\href {\doibase
  10.1016/j.vibspec.2008.01.014} {\bibfield  {journal} {\bibinfo  {journal}
  {Vibrational Spectroscopy}\ }\textbf {\bibinfo {volume} {47}},\ \bibinfo
  {pages} {53} (\bibinfo {year} {2008})}\BibitemShut {NoStop}%
\bibitem [{\citenamefont {Sato}\ \emph {et~al.}(2014)\citenamefont {Sato},
  \citenamefont {Aoki}, \citenamefont {Kino}, \citenamefont {Kuroe},
  \citenamefont {Sekine}, \citenamefont {Hase}, \citenamefont {Oka},
  \citenamefont {Ito},\ and\ \citenamefont {Eisaki}}]{sato2014raman}%
  \BibitemOpen
  \bibfield  {author} {\bibinfo {author} {\bibfnamefont {T.}~\bibnamefont
  {Sato}}, \bibinfo {author} {\bibfnamefont {K.}~\bibnamefont {Aoki}}, \bibinfo
  {author} {\bibfnamefont {R.}~\bibnamefont {Kino}}, \bibinfo {author}
  {\bibfnamefont {H.}~\bibnamefont {Kuroe}}, \bibinfo {author} {\bibfnamefont
  {T.}~\bibnamefont {Sekine}}, \bibinfo {author} {\bibfnamefont
  {M.}~\bibnamefont {Hase}}, \bibinfo {author} {\bibfnamefont {K.}~\bibnamefont
  {Oka}}, \bibinfo {author} {\bibfnamefont {T.}~\bibnamefont {Ito}}, \ and\
  \bibinfo {author} {\bibfnamefont {H.}~\bibnamefont {Eisaki}},\ }in\
  \href@noop {} {\emph {\bibinfo {booktitle} {Proceedings of the International
  Conference on Strongly Correlated Electron Systems (SCES2013)}}}\ (\bibinfo
  {year} {2014})\ p.\ \bibinfo {pages} {014035}\BibitemShut {NoStop}%
\bibitem [{\citenamefont {Mikhaylovskiy}\ \emph {et~al.}(2015)\citenamefont
  {Mikhaylovskiy}, \citenamefont {Hendry}, \citenamefont {Secchi},
  \citenamefont {Mentink}, \citenamefont {Eckstein}, \citenamefont {Wu},
  \citenamefont {Pisarev}, \citenamefont {Kruglyak}, \citenamefont
  {Katsnelson}, \citenamefont {Rasing},\ and\ \citenamefont
  {Kimel}}]{mikhaylovskiy2015ultrafast}%
  \BibitemOpen
  \bibfield  {author} {\bibinfo {author} {\bibfnamefont {R.~V.}\ \bibnamefont
  {Mikhaylovskiy}}, \bibinfo {author} {\bibfnamefont {E.}~\bibnamefont
  {Hendry}}, \bibinfo {author} {\bibfnamefont {A.}~\bibnamefont {Secchi}},
  \bibinfo {author} {\bibfnamefont {J.~H.}\ \bibnamefont {Mentink}}, \bibinfo
  {author} {\bibfnamefont {M.}~\bibnamefont {Eckstein}}, \bibinfo {author}
  {\bibfnamefont {A.}~\bibnamefont {Wu}}, \bibinfo {author} {\bibfnamefont
  {R.~V.}\ \bibnamefont {Pisarev}}, \bibinfo {author} {\bibfnamefont {V.~V.}\
  \bibnamefont {Kruglyak}}, \bibinfo {author} {\bibfnamefont {M.~I.}\
  \bibnamefont {Katsnelson}}, \bibinfo {author} {\bibfnamefont
  {T.}~\bibnamefont {Rasing}}, \ and\ \bibinfo {author} {\bibfnamefont {A.~V.}\
  \bibnamefont {Kimel}},\ }\href {\doibase 10.1038/ncomms9190} {\bibfield
  {journal} {\bibinfo  {journal} {Nature communications}\ }\textbf {\bibinfo
  {volume} {6}},\ \bibinfo {pages} {8190} (\bibinfo {year} {2015})}\BibitemShut
  {NoStop}%
\bibitem [{\citenamefont {Giacovazzo}(2002)}]{giacovazzo2002fundamentals}%
  \BibitemOpen
  \bibfield  {author} {\bibinfo {author} {\bibfnamefont {C.}~\bibnamefont
  {Giacovazzo}},\ }\href@noop {} {\emph {\bibinfo {title} {Fundamentals of
  crystallography}}},\ Vol.~\bibinfo {volume} {7}\ (\bibinfo  {publisher}
  {Oxford university press, USA},\ \bibinfo {year} {2002})\BibitemShut
  {NoStop}%
\bibitem [{\citenamefont {Hoppe}\ \emph {et~al.}(1989)\citenamefont {Hoppe},
  \citenamefont {Voigt}, \citenamefont {Glaum}, \citenamefont {Kissel},
  \citenamefont {M{\"u}ller},\ and\ \citenamefont {Bernet}}]{hoppe1989new}%
  \BibitemOpen
  \bibfield  {author} {\bibinfo {author} {\bibfnamefont {R.}~\bibnamefont
  {Hoppe}}, \bibinfo {author} {\bibfnamefont {S.}~\bibnamefont {Voigt}},
  \bibinfo {author} {\bibfnamefont {H.}~\bibnamefont {Glaum}}, \bibinfo
  {author} {\bibfnamefont {J.}~\bibnamefont {Kissel}}, \bibinfo {author}
  {\bibfnamefont {H.~P.}\ \bibnamefont {M{\"u}ller}}, \ and\ \bibinfo {author}
  {\bibfnamefont {K.}~\bibnamefont {Bernet}},\ }\href {\doibase
  10.1016/0022-5088(89)90411-6} {\bibfield  {journal} {\bibinfo  {journal}
  {Journal of the Less Common Metals}\ }\textbf {\bibinfo {volume} {156}},\
  \bibinfo {pages} {105} (\bibinfo {year} {1989})}\BibitemShut {NoStop}%
\bibitem [{\citenamefont {Tunell}\ \emph {et~al.}(1935)\citenamefont {Tunell},
  \citenamefont {Posnjak},\ and\ \citenamefont
  {Ksanda}}]{tunell1935geometrical}%
  \BibitemOpen
  \bibfield  {author} {\bibinfo {author} {\bibfnamefont {G.}~\bibnamefont
  {Tunell}}, \bibinfo {author} {\bibfnamefont {E.}~\bibnamefont {Posnjak}}, \
  and\ \bibinfo {author} {\bibfnamefont {C.~J.}\ \bibnamefont {Ksanda}},\
  }\href {\doibase 10.1524/zkri.1935.90.1.120} {\bibfield  {journal} {\bibinfo
  {journal} {Zeitschrift f{\"u}r Kristallographie-Crystalline Materials}\
  }\textbf {\bibinfo {volume} {90}},\ \bibinfo {pages} {120} (\bibinfo {year}
  {1935})}\BibitemShut {NoStop}%
\bibitem [{\citenamefont {Pisarev}\ \emph {et~al.}(2016)\citenamefont
  {Pisarev}, \citenamefont {Prosnikov}, \citenamefont {Davydov}, \citenamefont
  {Smirnov}, \citenamefont {Roginskii}, \citenamefont {Boldyrev}, \citenamefont
  {Molchanova}, \citenamefont {Popova}, \citenamefont {Smirnov},\ and\
  \citenamefont {Kazimirov}}]{pisarev2016lattice}%
  \BibitemOpen
  \bibfield  {author} {\bibinfo {author} {\bibfnamefont {R.~V.}\ \bibnamefont
  {Pisarev}}, \bibinfo {author} {\bibfnamefont {M.~A.}\ \bibnamefont
  {Prosnikov}}, \bibinfo {author} {\bibfnamefont {V.~Y.}\ \bibnamefont
  {Davydov}}, \bibinfo {author} {\bibfnamefont {A.~N.}\ \bibnamefont
  {Smirnov}}, \bibinfo {author} {\bibfnamefont {E.~M.}\ \bibnamefont
  {Roginskii}}, \bibinfo {author} {\bibfnamefont {K.~N.}\ \bibnamefont
  {Boldyrev}}, \bibinfo {author} {\bibfnamefont {A.~D.}\ \bibnamefont
  {Molchanova}}, \bibinfo {author} {\bibfnamefont {M.~N.}\ \bibnamefont
  {Popova}}, \bibinfo {author} {\bibfnamefont {M.~B.}\ \bibnamefont {Smirnov}},
  \ and\ \bibinfo {author} {\bibfnamefont {V.~Y.}\ \bibnamefont {Kazimirov}},\
  }\href {\doibase 10.1103/PhysRevB.93.134306} {\bibfield  {journal} {\bibinfo
  {journal} {Phys. Rev. B}\ }\textbf {\bibinfo {volume} {93}},\ \bibinfo
  {pages} {134306} (\bibinfo {year} {2016})}\BibitemShut {NoStop}%
\bibitem [{\citenamefont {Abdullaev}\ and\ \citenamefont
  {Mamedov}(1981)}]{abdullaev1981refined}%
  \BibitemOpen
  \bibfield  {author} {\bibinfo {author} {\bibfnamefont {G.~K.}\ \bibnamefont
  {Abdullaev}}\ and\ \bibinfo {author} {\bibfnamefont {K.~S.}\ \bibnamefont
  {Mamedov}},\ }\href {\doibase 10.1007/BF00784113} {\bibfield  {journal}
  {\bibinfo  {journal} {Journal of Structural Chemistry}\ }\textbf {\bibinfo
  {volume} {22}},\ \bibinfo {pages} {637} (\bibinfo {year} {1981})}\BibitemShut
  {NoStop}%
\bibitem [{\citenamefont {Bassi}\ \emph {et~al.}(1996)\citenamefont {Bassi},
  \citenamefont {Camagni}, \citenamefont {Rolli}, \citenamefont {Samoggia},
  \citenamefont {Parmigiani}, \citenamefont {Dhalenne},\ and\ \citenamefont
  {Revcolevschi}}]{bassi1996optical}%
  \BibitemOpen
  \bibfield  {author} {\bibinfo {author} {\bibfnamefont {M.}~\bibnamefont
  {Bassi}}, \bibinfo {author} {\bibfnamefont {P.}~\bibnamefont {Camagni}},
  \bibinfo {author} {\bibfnamefont {R.}~\bibnamefont {Rolli}}, \bibinfo
  {author} {\bibfnamefont {G.}~\bibnamefont {Samoggia}}, \bibinfo {author}
  {\bibfnamefont {F.}~\bibnamefont {Parmigiani}}, \bibinfo {author}
  {\bibfnamefont {G.}~\bibnamefont {Dhalenne}}, \ and\ \bibinfo {author}
  {\bibfnamefont {A.}~\bibnamefont {Revcolevschi}},\ }\href {\doibase
  10.1103/PhysRevB.54.R11030} {\bibfield  {journal} {\bibinfo  {journal} {Phys.
  Rev. B}\ }\textbf {\bibinfo {volume} {54}},\ \bibinfo {pages} {R11030}
  (\bibinfo {year} {1996})}\BibitemShut {NoStop}%
\bibitem [{\citenamefont {Popova}\ \emph {et~al.}(1996)\citenamefont {Popova},
  \citenamefont {Sushkov}, \citenamefont {Golubchik}, \citenamefont
  {Vasil'ev},\ and\ \citenamefont {Leonyuk}}]{popova1996optical}%
  \BibitemOpen
  \bibfield  {author} {\bibinfo {author} {\bibfnamefont {M.~N.}\ \bibnamefont
  {Popova}}, \bibinfo {author} {\bibfnamefont {A.~B.}\ \bibnamefont {Sushkov}},
  \bibinfo {author} {\bibfnamefont {S.~A.}\ \bibnamefont {Golubchik}}, \bibinfo
  {author} {\bibfnamefont {A.~N.}\ \bibnamefont {Vasil'ev}}, \ and\ \bibinfo
  {author} {\bibfnamefont {L.~I.}\ \bibnamefont {Leonyuk}},\ }\href@noop {}
  {\bibfield  {journal} {\bibinfo  {journal} {JETP}\ }\textbf {\bibinfo
  {volume} {83}},\ \bibinfo {pages} {1227} (\bibinfo {year}
  {1996})}\BibitemShut {NoStop}%
\bibitem [{\citenamefont {van Loosdrecht}(1998)}]{van1998optical}%
  \BibitemOpen
  \bibfield  {author} {\bibinfo {author} {\bibfnamefont {P.~H.~M.}\
  \bibnamefont {van Loosdrecht}},\ }in\ \href {\doibase
  10.4028/www.scientific.net/SSP.61-62.19} {\emph {\bibinfo {booktitle}
  {Contemporary Studies in Condensed Matter Physics}}},\ \bibinfo {series}
  {Solid State Phenomena}, Vol.~\bibinfo {volume} {61}\ (\bibinfo  {publisher}
  {Trans Tech Publications},\ \bibinfo {year} {1998})\ pp.\ \bibinfo {pages}
  {19--26}\BibitemShut {NoStop}%
\bibitem [{\citenamefont {Burns}(1993)}]{burns1993mineralogical}%
  \BibitemOpen
  \bibfield  {author} {\bibinfo {author} {\bibfnamefont {R.~G.}\ \bibnamefont
  {Burns}},\ }\href@noop {} {\emph {\bibinfo {title} {Mineralogical
  applications of crystal field theory}}},\ Vol.~\bibinfo {volume} {5}\
  (\bibinfo  {publisher} {Cambridge University Press},\ \bibinfo {year}
  {1993})\BibitemShut {NoStop}%
\bibitem [{\citenamefont {Ray}(2001)}]{ray2001preparation}%
  \BibitemOpen
  \bibfield  {author} {\bibinfo {author} {\bibfnamefont {S.~C.}\ \bibnamefont
  {Ray}},\ }\href@noop {} {\bibfield  {journal} {\bibinfo  {journal} {Solar
  energy materials and solar cells}\ }\textbf {\bibinfo {volume} {68}},\
  \bibinfo {pages} {307} (\bibinfo {year} {2001})}\BibitemShut {NoStop}%
\bibitem [{\citenamefont {Tahir}\ and\ \citenamefont
  {Tougaard}(2012)}]{tahir2012electronic}%
  \BibitemOpen
  \bibfield  {author} {\bibinfo {author} {\bibfnamefont {D.}~\bibnamefont
  {Tahir}}\ and\ \bibinfo {author} {\bibfnamefont {S.}~\bibnamefont
  {Tougaard}},\ }\href {\doibase 10.1088/0953-8984/24/17/175002} {\bibfield
  {journal} {\bibinfo  {journal} {Journal of Physics: Condensed Matter}\
  }\textbf {\bibinfo {volume} {24}},\ \bibinfo {pages} {175002} (\bibinfo
  {year} {2012})}\BibitemShut {NoStop}%
\bibitem [{\citenamefont {Ching}\ \emph {et~al.}(1989)\citenamefont {Ching},
  \citenamefont {Xu},\ and\ \citenamefont {Wong}}]{ching1989ground}%
  \BibitemOpen
  \bibfield  {author} {\bibinfo {author} {\bibfnamefont {W.~Y.}\ \bibnamefont
  {Ching}}, \bibinfo {author} {\bibfnamefont {Y.-N.}\ \bibnamefont {Xu}}, \
  and\ \bibinfo {author} {\bibfnamefont {K.~W.}\ \bibnamefont {Wong}},\ }\href
  {\doibase 10.1103/PhysRevB.40.7684} {\bibfield  {journal} {\bibinfo
  {journal} {Phys. Rev. B}\ }\textbf {\bibinfo {volume} {40}},\ \bibinfo
  {pages} {7684} (\bibinfo {year} {1989})}\BibitemShut {NoStop}%
\bibitem [{\citenamefont {Kuroe}\ \emph {et~al.}(1994)\citenamefont {Kuroe},
  \citenamefont {Sekine}, \citenamefont {Hase}, \citenamefont {Sasago},
  \citenamefont {Uchinokura}, \citenamefont {Kojima}, \citenamefont {Tanaka},\
  and\ \citenamefont {Shibuya}}]{kuroe1994raman}%
  \BibitemOpen
  \bibfield  {author} {\bibinfo {author} {\bibfnamefont {H.}~\bibnamefont
  {Kuroe}}, \bibinfo {author} {\bibfnamefont {T.}~\bibnamefont {Sekine}},
  \bibinfo {author} {\bibfnamefont {M.}~\bibnamefont {Hase}}, \bibinfo {author}
  {\bibfnamefont {Y.}~\bibnamefont {Sasago}}, \bibinfo {author} {\bibfnamefont
  {K.}~\bibnamefont {Uchinokura}}, \bibinfo {author} {\bibfnamefont
  {H.}~\bibnamefont {Kojima}}, \bibinfo {author} {\bibfnamefont
  {I.}~\bibnamefont {Tanaka}}, \ and\ \bibinfo {author} {\bibfnamefont
  {Y.}~\bibnamefont {Shibuya}},\ }\href {\doibase 10.1103/PhysRevB.50.16468}
  {\bibfield  {journal} {\bibinfo  {journal} {Phys. Rev. B}\ }\textbf {\bibinfo
  {volume} {50}},\ \bibinfo {pages} {16468} (\bibinfo {year}
  {1994})}\BibitemShut {NoStop}%
\bibitem [{\citenamefont {Damascelli}\ \emph {et~al.}(1997)\citenamefont
  {Damascelli}, \citenamefont {van~der Marel}, \citenamefont {Parmigiani},
  \citenamefont {Dhalenne},\ and\ \citenamefont
  {Revcolevschi}}]{damascelli1997infrared}%
  \BibitemOpen
  \bibfield  {author} {\bibinfo {author} {\bibfnamefont {A.}~\bibnamefont
  {Damascelli}}, \bibinfo {author} {\bibfnamefont {D.}~\bibnamefont {van~der
  Marel}}, \bibinfo {author} {\bibfnamefont {F.}~\bibnamefont {Parmigiani}},
  \bibinfo {author} {\bibfnamefont {G.}~\bibnamefont {Dhalenne}}, \ and\
  \bibinfo {author} {\bibfnamefont {A.}~\bibnamefont {Revcolevschi}},\ }\href
  {\doibase 10.1103/PhysRevB.56.R11373} {\bibfield  {journal} {\bibinfo
  {journal} {Phys. Rev. B}\ }\textbf {\bibinfo {volume} {56}},\ \bibinfo
  {pages} {R11373} (\bibinfo {year} {1997})}\BibitemShut {NoStop}%
\bibitem [{\citenamefont {Popova}\ \emph {et~al.}(1998)\citenamefont {Popova},
  \citenamefont {Sushkov}, \citenamefont {Golubchik}, \citenamefont
  {Vasil'ev},\ and\ \citenamefont {Leonyuk}}]{popova1998folded}%
  \BibitemOpen
  \bibfield  {author} {\bibinfo {author} {\bibfnamefont {M.~N.}\ \bibnamefont
  {Popova}}, \bibinfo {author} {\bibfnamefont {A.~B.}\ \bibnamefont {Sushkov}},
  \bibinfo {author} {\bibfnamefont {S.~A.}\ \bibnamefont {Golubchik}}, \bibinfo
  {author} {\bibfnamefont {A.~N.}\ \bibnamefont {Vasil'ev}}, \ and\ \bibinfo
  {author} {\bibfnamefont {L.~I.}\ \bibnamefont {Leonyuk}},\ }\href {\doibase
  10.1103/PhysRevB.57.5040} {\bibfield  {journal} {\bibinfo  {journal} {Phys.
  Rev. B}\ }\textbf {\bibinfo {volume} {57}},\ \bibinfo {pages} {5040}
  (\bibinfo {year} {1998})}\BibitemShut {NoStop}%
\end{thebibliography}%


%

\end{document}